\documentclass[aip,cha, reprint,showpacs,showkeys]{revtex4-1}
\usepackage{amssymb, dcolumn,  graphicx, latexsym, amsfonts, bm}

\begin{document}

\title{Non-modal kinetic theory of the stability of the compressed-sheared plasma 
flows generated by the inhomogeneous microscale turbulence in the tokamak edge plasma }

\author{V. S. Mikhailenko}\email[E-mail:]{vsmikhailenko@pusan.ac.kr}
\affiliation{Plasma Research Center, Pusan National University, Busan 46241, South Korea}
\author{V. V. Mikhailenko}\email[E-mail: ]{vladimir@pusan.ac.kr}
\affiliation{BK21 FOUR Information Technology, Pusan National University, Busan 46241, 
South Korea}
\author{Hae June Lee}\email[E-mail: ]{haejune@pusan.ac.kr}
\affiliation{Department of Electrical Engineering, Pusan National University, Busan 46241, South 
Korea}
\date{\today}

\begin{abstract}
A nonmodal kinetic theory of the stability of the two-dimensional 
compressed-sheared mesoscale plasma flows, generated by the radially inhomogeneous electrostatic 
ion cyclotron parametric microturbulence in the pedestal plasma with a sheared poloidal flow, is 
developed. This theory reveals that the separate spatially uniform Fourier modes of the 
electrostatic responses of the ions and of the electrons on the mesoscale convective flows 
are determined only in the 
frames of references moved with velocities of the ion and electron convective flows. 
In the laboratory frame, these modes are observed as the compressed-sheared modes with time 
dependent wave numbers. The integral equation, which governs the separate Fourier mode of the 
electrostatic potential of the plasma species responses on the mesoscale convective flows, 
is derived. In this equation, the effects of the compressing and shearing of the convective flows 
are revealed as the time dependence of the finite ion Larmor radius effect. The solution of this 
equation for the kinetic drift instability displays the nonmodal transformation of the 
potential to the zero frequency cell-like perturbation when time elapsed. 
\end{abstract}

\maketitle

 \section{Introduction}\label{sec1}

The linear theory of the interaction of the fast waves (FW) with tokamak plasma predicts 
\cite{Ono, Chiu} that the injection of FW may be the efficient method for the 
electron heating and current drive to aid in steady-state non-inductive tokamak operation. 
These predictions has been confirmed for the propagation and absorption 
of FWs in the hot core tokamak plasma bounded by the last closed flux surface (LCFS) 
on numerous tokamak devices over the past half century. 
These experiments, however, demonstrated that the efficiency of the FW heating and current drive reduces by the FW power lost, which occurs in the near-antenna layer of the cold low 
density scrape-off layer (SOL) tokamak plasma. The FW heating experiments on the National 
Spherical Torus eXperiment (NSTX) showed\cite{Perkins, Perkins1} that around 30\% to more 
than 60\% of the FW energy lost directly in the SOL. The bursts of the ions 
with energy above 20 keV, experimentally observed\cite{Pace} in SOL following FW injection, 
and the development of the parametric instabilities in SOL \cite{Wilson}, predicted earlier 
theoretically\cite{Porkolab1,Porkolab2}, were considered as the main channels of the FW 
absorption in SOL. The development of the ion cyclotron (IC) quasimode decay instability was 
considered\cite{Pace,Wilson} as the main nonlinear process responsible  
for the absorption of the FW power in SOL plasma and of the anomalous heating of ions in SOL.
The analysis of the turbulent heating of ions by the IC parametric turbulence, 
powered by the IC quasimode decay instability, was given in Ref.\cite{Mikhailenko}
on the base of the numerical solution of the dispersion equation for the IC parametric 
instabilities driven by FW, and on the base of quasilinear theory for ion distribution function, 
which accounted for the interaction of ions with IC parametric turbulence 
powered by the IC quasimode decay instability. The derived estimates for the turbulent ion heating 
rates revealed that the absorption of the FW energy by ions in SOL is a weak effect, which 
provides only negligibly small heating of cold ions in SOL and can not be responsible for the 
observed generation of the high energy ions. 
 
The FW power loss in SOL was investigated in Refs. \cite{Bertelli, Bertelli1} by using the numerical 
full wave simulation code AORSA (all-orders spectral algorithm)\cite{Jaeger, Green}, in which the 
edge plasma beyond LCFS is included in the solution domain. This simulation displays that the 
dominant loss of the FW power occurs in the edge of the tokamak plasma, where the 
plasma density and the amplitude of the FW field strongly change on the radial  
intermediate spatial scale (mesoscale) between the 
macroscale of the spatial inhomogeneity length of the FW field in the bulk of the tokamak plasma, 
and the microscale commensurable with the wavelength of the parametric instabilities
of the IC perturbations. The theory of the microscale parametric turbulence of the inhomogeneous 
plasma driven by the strong inhomogeneous on the mesoscale FW, was developed in Ref.
\cite{Mikhailenko1}. This theory reveals the effect of the formation of the radial and poloidal 
convective flows of such a plasma caused by the mesoscale spatial inhomogeneity of the microscale 
IC or drift turbulence. The radial and poloidal convective flow velocity components are proportional 
to the gradient of the spectral intensity of the electric field of the microturbulence. This result 
gives the possible explanation of the FW heating experiment on the National Spherical 
Torus eXperiment (NSTX)\cite{Perkins,Perkins1,Hosea}, where it was found that a significant part of 
the FW power loss occurs due to the anomalous convective flow of the collisionless dense hot plasma 
from the tokamak edge to the cold low density SOL plasma.

The pedestal region, where plasma density profile has largest radial gradient, is the most 
preferable region in tokamaks for the development of the convective flows, driven by the 
spatially inhomogeneous microturbulence. The inherent 
component of the pedestal plasma is the sheared poloidal flow, in which the drift type 
instabilities, responsible for the anomalous transport of plasma, are suppressed 
when their growth rates are less than the flow velocity shearing rate\cite{Burrell}. 
It was proved in Refs. \cite{Mikhailenko2,Mikhailenko4,Mikhailenko5} that the basic point in 
understanding the processes of the instabilities and turbulence evolution in plasma 
sheared flows is the proper treatment of the persistent deformation 
of the perturbations by the sheared flows. This effect, which is completely ignored in 
the canonical stability theory, where the perturbations are considered having a static 
structure of a plane wave $\sim \exp \left(i\mathbf{k}\cdot\mathbf{r}-i\omega t\right)$,
is involved in the nonmodal kinetic theory, developed in Refs. 
\cite{Mikhailenko2,Mikhailenko4,Mikhailenko5},  
grounded on the methodology of the sheared modes. It was found\cite{Mikhailenko2,Mikhailenko4}
that in the sheared flow, the separate spatial Fourier mode with a static spatial structure 
$\sim \exp \left(ik_{x}x+ik_{y}y+ik_{z}z\right)$ 
can be determined only in the frame convected with a sheared flow. In the 
laboratory frame, this mode is observed as the sheared mode with time dependent 
structure resulted from the continuous distortion with time the perturbation by the 
sheared flow. This distortion grows with time and forms a time-dependent nonmodal process which 
is investigated as the initial value problem.

In Ref.\cite{Mikhailenko3}, the Vlasov equations, which govern the ion and electron mesoscale 
convective flows with radially inhomogeneous flow velocities in the poloidal sheared flow were 
derived. These equations predict the generation of the sheared poloidal convective flow 
and of the radial compressed flow with radial flow velocity gradient. The hydrodynamic theory 
of the mesoscale convective flows, derived as the moments of the obtained Vlasov equation, 
reveals the radial compressed convective flow as the dominant factor in the 
formation of the steep pedestal density profile with density gradient exponentially 
growing with time. The focus of this paper is the development of the nonmodal kinetic 
theory of the stability of the two-dimensional compressed-sheared convective flow.
In Sec. \ref{sec2}, we present basic equations and their transformations. 
In Sec. \ref{sec3}, we develop the nonmodal approach 
to the kinetic theory of the compressed-sheared convective flows. In this theory we derived new  
spatial reference coordinates in which the distribution functions of the unperturbed convective 
sheared-compressed flows is stationary. The stability of such a distribution functions of the 
convected plasma species against the development of the short scale instabilities is given 
in Sec. \ref{sec4} employing the developed compressed-sheared modes approach. The Conclusions are 
given in Sec.  \ref{sec5}.

 \section{Basic equations and transformations}\label{sec2} 
Our theory is based on the Vlasov-Poisson system in a slab geometry approximation 
where $x, y, z$ directions are viewed as corresponding to the radial, poloidal and 
toroidal directions, respectively, of the toroidal coordinate system. Within this 
approximation, the Vlasov equation for the velocity distribution function $F_{\alpha}$ 
of the poloidal sheared flow of $\alpha$ plasma 
species ($\alpha=i$ for ions and $\alpha=e$ for electrons) in the FW field with coordinates 
$\mathbf{r}=\left(x, y, z\right)$ has a form
\begin{eqnarray}
&\displaystyle \frac{\partial F_{\alpha}\left(\mathbf{v}, \mathbf{r}, t\right)}
{\partial t}+\mathbf{v}\frac{\partial F_{\alpha}\left(\mathbf{v}, \mathbf{r}, t\right)}
{\partial\mathbf{r}}
\nonumber 
\\ 
&\displaystyle
+\frac{e_{\alpha}}{m_{\alpha}}\left(\mathbf{E}_{0x}\left(x\right)
+\mathbf{E}_{1}\left(x, t\right)+\tilde{\mathbf{E}}\left(\mathbf{r}, t\right)\right.
\nonumber 
\\ 
&\displaystyle
\left.+\frac{1}{c}\left[\mathbf{v}\times\left(\mathbf{B}_{0}+
\mathbf{B}_{1}\left(\mathbf{r}, t\right)\right)\right]\right)\frac{\partial
F_{\alpha}\left(\mathbf{v}, \mathbf{r}, t\right)}{\partial\mathbf{v}}=0.
\label{1}
\end{eqnarray}
This equation contains the inhomogeneous radial electric field $\mathbf{E}_{0x}\left(x\right)$, 
which governs the basic poloidal sheared flow, the FW electric field $\mathbf{E}_{1}
\left(x, t\right)$, the electric field $\tilde{\mathbf{E}}\left(\mathbf{r}, t\right)$ of 
the self-consistent plasma response on FW, the uniform plasma-confining magnetic field 
$\mathbf{B}_{0}$ directed along coordinate $z$, and FW magnetic field 
$\mathbf{B}_{1}\left(\mathbf{r}, t\right)$. 
For the edge layer of the tokamak plasma, this equation contains two disparate spatial 
inhomogeneity lengths, which are introduced by the FW field and by plasma parameters. 
In the edge plasma, the spatial inhomogeneity of $\mathbf{E}_{0x}\left(x\right)$ and of FW 
fields are commensurable with a spatial inhomogeneity length of the pedestal plasma density. These 
mesoscale spatial inhomogeneity lengths on order of the pedestal width are much less than the 
the inhomogeneity scale lengths of FW and of the plasma parameters in the plasma core, 
but are much larger than the radial wavelengths of the IC parametric and drift microturbulence. 
Electric field $\tilde{\mathbf{E}}\left(\mathbf{r}, t\right)$, being the microscale responc of the 
inhomogeneous plasma on the inhomogeneous FW and $\mathbf{E}_{0x}\left(x\right)$ fields, 
contains micro- and meso- spatial scales. Our theory of the mesoscale plasma evolution 
caused by the mesoscale inhomogeneities of the microturbulence, involves the treatments on 
the micro- and mesoscales. In our theory\cite{Mikhailenko3}, we introduced jointly with variables 
$\mathbf{r}=\left(x, y, z\right)$ and time $t$ for the microscale fast variations on time 
of the order of the FW period or of the period of the IC microturbulence, 
the slow time $T=\varepsilon t$, and the slow spatial variables 
$X=\varepsilon x$, $Y=\varepsilon y$, where the dimensionless parameter $\varepsilon\ll 1$, 
for the description of the slow evolutionary mesoscale processes 
in the pedestal region. With these microscale and mesoscale 
variables, electric field $\mathbf{E}_{0x}$ depends only on $X$. The FW fields 
$\mathbf{E}_{1}$, $\mathbf{B}_{1}$, determined as 
\begin{eqnarray}
&\displaystyle \mathbf{E}_{1}\left(X, t\right)=
\mathbf{E}_{1x}\left(X\right)\cos \omega_{0}t+ \mathbf{E}_{1y}\left(X\right)\sin \omega_{0}t,
\label{2}
\end{eqnarray}
\begin{eqnarray}
&\displaystyle 
\mathbf{B}_{1}\left(X, t\right)=\frac{c}{\omega_{0}}\frac{dE_{1y}\left(X\right)}{d X}
\cos \omega_{0}t\,\mathbf{e}_{z}.
\label{3}
\end{eqnarray}
depend on slow mesoscale $X$ and fast time $t$. The electric field $\tilde{\mathbf{E}}
\left(\mathbf{r}, X, t\right)$ depends on the spatial micro- and mesoscale variables and on the 
fast time. This field is determined by the Poisson equation 
\begin{eqnarray}
&\displaystyle 
\nabla\cdot \tilde{\mathbf{E}}\left(\mathbf{r}, X, t\right)=
4\pi\sum_{\alpha=i,e} e_{\alpha}\int f_{\alpha}\left(\mathbf{v}, \mathbf{r}, X, t \right)d{\bf v}, 
\label{4}
\end{eqnarray}
in which $f_{\alpha}$ is the fluctuating part of the distribution function 
$F_{\alpha}$, $f_{\alpha}=F_{\alpha}-F_{0\alpha}$, where $F_{0\alpha}$ is the equilibrium 
distribution function. 

It is obvious that it is not possible to apply directly to the Vlasov-Poisson system (\ref{1}), 
(\ref{4}) with spatially inhomogeneous oscillating FW fields the methods of the 
solutions known for the investigations of the stability of a plasma in static equilibrium. 
Any microscale perturbations of the ion and electron densities are convected by FW field with 
inhomogeneous on the mesoscale velocities, different for ions 
and electrons, and oscillating with frequency of the FW field. It was found in 
Ref.\cite{Mikhailenko1} that the spatially inhomogeneous FW field may be excluded from the Vlasov 
equation (\ref{1}) by the transformation of the velocity $\mathbf{v}$ and the 
position $\mathbf{r}=\left(x, y, z\right)$ variables of the Vlasov equation (\ref{1}) to new 
velocity $\mathbf{v}_{i}$ and position $\mathbf{r}_{i}=\left(x_{i}, y_{i}, z\right)$ variables 
determined in the convected reference flow, which moves relative to the laboratory frame
with the velocity $\mathbf{V}_{i}\left(X_{i}, t\right)$ of ion in the FW field, 
given by the equation
\begin{eqnarray}
&\displaystyle 
\frac{d\mathbf{V}_{i}\left(X_{i}, t\right)}{dt}
=\frac{e_{i}}{m_{\alpha}}\left(\mathbf{E}_{1}\left(X_{i}+
\varepsilon R_{ix}\left(X_{i}, t\right), t\right)
\right.
\nonumber 
\\
&\displaystyle
\left.+\frac{1}{c}\left[\mathbf{V}_{i}\times\mathbf{B}_{0}\right]+\frac{1}{c}
\left[\mathbf{V}_{i}\times\mathbf{B}_{1}\left(X_{i}+\varepsilon R_{ix}
\left(X_{i}, t\right), t\right)\right]\right),
\label{5}
\end{eqnarray}
with initial value $\mathbf{V}_{i}\left(X_{i}, t=t_{0}=0\right)=0$. The Vlasov equation 
(\ref{1}) with new variables $\mathbf{v}_{i}$, $X_{i}$,  
contains the electric FW field only in terms on the order of $|R_{i}/L_{E}|\ll 1$, 
where $|R_{i}|$ is the amplitude of the ion displacement in the spatially 
inhomogeneous FW field with spatial inhomogeneity scale length $L_{E}$. These terms are 
negligibly small\cite{Mikhailenko1} for the conditions of the FW tokamak plasma heating 
and may be neglected. Without these terms, the Vlasov equation in the frame 
convected with velocity $\mathbf{V}_{i}\left(X_{i}, t\right)$ has a form as for 
a static equilibria without the external FW field. That equation for ions,
\begin{eqnarray}
&\displaystyle 
\frac{\partial F_{i}\left(\mathbf{v}_{i}, \mathbf{r}_{i}, X_{i}, t\right)}
{\partial t}+ \mathbf{v}_{i}\frac{\partial F_{i}} {\partial
\mathbf{r}_{i}}+\frac{e_{i}}{m_{i}c}\left[\mathbf{v}_{i}\times\mathbf{B}_{0}\right]
\frac{\partial F_{i}}{\partial\mathbf{v}_{i}}
\nonumber 
\\ 
&\displaystyle
+\frac{e_{i}}{m_{i}}\tilde{\mathbf{E}}_{i}\left(\mathbf{r}_{i}, X_{i}, t\right)
\frac{\partial F_{i}\left(\mathbf{v}_{i}, \mathbf{r}_{i}, X_{i}, 
t\right)}{\partial \mathbf{v}_{i}} =0,
\label{6}
\end{eqnarray}
and similar equation for electrons, determined in the electron reference flow, and the 
Poisson equation for the electric field 
\begin{eqnarray}
&\displaystyle 
\nabla\cdot \tilde{\mathbf{E}}_{i}\left(\mathbf{r}_{i}, X_{i}, t\right)=
4\pi\sum_{\alpha=i,e} e_{\alpha}\int f_{\alpha}\left(\mathbf{v}_{\alpha}, 
\mathbf{r}_{\alpha}, X_{\alpha}, t \right)d\mathbf{v}_{\alpha}, 
\label{7}
\end{eqnarray}
determined in the ion reference flow, compose the system of equations for the investigation 
of the microscale turbulence in FW field. The mesoscale variables $X_{i}$ and $X_{e}$ 
are presented in this system as parameters. 

At the time above the inverse growth rate of the IC instabilities, 
$t\gg \gamma^{-1}\left(\mathbf{k}\right)> |\omega^{-1}\left(\mathbf{k}\right)|$ the microscale IC 
turbulence attains the steady state. At this state the electric field $\tilde{\mathbf{E}}_{i}$ 
of the electrostatic two dimensional IC parametric microturbulence, directed almost across the 
magnetic field $\mathbf{B}_{0}$, may be presented in the 
ion reference flow in the form
\begin{eqnarray}
&\displaystyle 
\tilde{\mathbf{E}}_{i}\left(\mathbf{r}_{i}, X_{i}, t\right)=
\sum \limits_{n}\tilde{\mathbf{E}}_{i}\left(\mathbf{r}_{i}, X_{i}, n, t\right)
\nonumber\\ 
& \displaystyle
= \frac{1}{\left(2\pi\right)^{2}}
\sum \limits_{n}\frac{1}{2}\int d\mathbf{k}\left[\tilde{\mathbf{E}}_{i}
\left(\mathbf{k}, X_{i}, n\right)
e^{i\psi_{n}\left(\mathbf{r}_{i}, X_{i}, t\right)}\right.
\nonumber\\ 
& \displaystyle
\left.+\tilde{\mathbf{E}}_{i}^{\ast}\left(\mathbf{k}, X_{i}, n\right)
e^{-i\psi_{n}\left(\mathbf{r}_{i}, X_{i}, t\right)} \right], 
\label{8}
\end{eqnarray}
where
\begin{eqnarray}
&\displaystyle
\psi_{n}\left(\mathbf{r}_{i}, X_{i}, t\right)= -i\omega_{n}\left(\mathbf{k}\right)t+ i\mathbf{k}
\mathbf{r}_{i}+i\theta\left(\mathbf{k}\right),
\label{9}
\end{eqnarray}
i. e. as a linear superposition of the electric fields of IC perturbations with 
frequencies $\omega_{n}\left(\mathbf{k}\right)=n\omega_{ci}+\delta\omega_{n}
\left(\mathbf{k}\right)$ with wave vectors $\mathbf{k}$ directed across the magnetic field 
and with mesoscale position dependent amplitudes $\tilde{\mathbf{E}}_{i}\left(\mathbf{k}, X_{i}, n
\right)$ and with phases fast changed with time on the microscales. 
Reality of $\tilde{\mathbf{E}}_{i}\left(\mathbf{r}_{i}, X_{i}, t\right)$
is insured without introducing negative frequencies by the addition of the complex conjugate 
terms with amplitudes $\tilde{\mathbf{E}}_{i}^{\ast}\left(\mathbf{k}, X_{i}, n\right)$. 
The integration over $\mathbf{k}$ is performed over wave vectors of the 
linearly unstable IC perturbations, and $\theta\left(\mathbf{k}\right)$ 
is their initial phase. In the electron frame, oscillating relative to the ion frame, 
this electric field is determined by Eq. (\ref{8}) with species subscript $i$ 
changed on $e$. The electric field 
$\tilde{\mathbf{E}}_{e}\left(\mathbf{r}_{e}, X_{e}, t\right)$ in variables 
$\mathbf{r}_{i}, X_{i}$ has a form
\begin{eqnarray}
&\displaystyle 
\tilde{\mathbf{E}}_{e}\left(\mathbf{r}_{e}, X_{e}, t\right)=\sum \limits_{n}\tilde{\mathbf{E}}_{e}
\left(\mathbf{r}_{e}, X_{e}, n, t\right)
\nonumber\\ 
& \displaystyle
=\frac{1}{\left(2\pi\right)^{2}}
\sum \limits_{n}\frac{1}{2}\int d\mathbf{k}\left[\tilde{\mathbf{E}}_{i}\left(\mathbf{k}, 
X_{i}, n\right)
\sum\limits_{p=-\infty}^{\infty}J_{p}\left(a_{ie}\right)\right.
\nonumber\\ 
& \displaystyle
\left.\times e^{i\psi_{n}\left(\mathbf{r}_{i}, X_{i}, 
t\right)-ip\left(\omega_{0}t+\delta_{ie}\left(\mathbf{k}, X_{i}\right)\right)}\right.
\nonumber\\ 
& \displaystyle
\left.+\tilde{\mathbf{E}}_{i}^{\ast}\left(\mathbf{k}, X_{i}, n\right)
\sum\limits_{p=-\infty}^{\infty}J_{p}\left(a_{ie}\right)\right.
\nonumber\\ 
& \displaystyle
\left.\times e^{-i\psi_{n}\left(\mathbf{r}_{i}, 
X_{i}, t\right)+ip\left(\omega_{0}t+\delta_{ie}\left(\mathbf{k}, X_{i}\right)\right)}\right],
\label{10}
\end{eqnarray}
where $J_{p}\left(a_{ie}\right)$ is the first kind Bessel 
function of order $p$ with argument $a_{ie}$. Parameters $a_{ie}\sim k\xi_{ie}$, 
where $\xi_{ie}$ is the amplitude of the relative displacement of electrons 
relative to ions in FW field, and $\delta_{ie}$ were determined in Ref.\cite{Mikhailenko}.

 \section{The kinetic theory of the mesoscale compressed-sheared convective flows}\label{sec3} 
For the investigation on the slow time scale $T$ the mesoscale evolution of the poloidal 
plasma sheared flow with microscale turbulence not suppressed by the sheared flow, we 
transform the Vlasov-Poisson system from the microscale to the mesoscale variables 
using the relations $x_{i}=\frac{1}{\varepsilon}X_{i}$,  $y_{i}=\frac{1}{\varepsilon}Y_{i}$ and 
$t=\frac{1}{\varepsilon}T$ variables. With mesoscale variables Eq. (\ref{6}) becomes
\begin{eqnarray}
&\displaystyle 
\varepsilon\frac{\partial F_{i}\left(\mathbf{v}_{i}, X_{i}, Y_{i}, T, \varepsilon\right)}
{\partial T}+ \varepsilon v_{ix}\frac{\partial F_{i}} {\partial
X_{i}}+ \varepsilon v_{iy}\frac{\partial F_{i}} {\partial
Y_{i}} 
\nonumber 
\\ 
&\displaystyle
+\frac{e_{i}}{m_{i}c}\left[\mathbf{v}_{i}\times\mathbf{B}_{0}\right]
\frac{\partial F_{i}}{\partial\mathbf{v}_{i}}
\nonumber 
\\ 
&\displaystyle
+\frac{e_{i}}{m_{i}}\tilde{\mathbf{E}}_{i}\left(X_{i}, Y_{i}, T, \varepsilon\right)
\frac{\partial F_{i}}{\partial \mathbf{v}_{i}} =0,
\label{11}
\end{eqnarray}
The electric field $\tilde{\mathbf{E}}_{i}\left(X_{i}, Y_{i}, T, \varepsilon\right)$ 
is determined by Eq. (\ref{8}), in which phase $\psi_{n}$ is determined in the form
\begin{eqnarray}
&\displaystyle
\psi_{n}\left( X_{i}, Y_{i}, T, \varepsilon\right)
\nonumber\\ 
& \displaystyle
= -i\frac{1}{\varepsilon}\left(\omega_{n}
\left(\mathbf{k}\right)T- ik_{x}X_{i}- ik_{y}Y_{i}\right)+i\theta\left(\mathbf{k}\right).
\label{12}
\end{eqnarray}
The similar transformations should be performed for the electron Vlasov equation.
On the nonlinear stage of the IC parametric microturbulence evolution at the time above 
the inverse growth rate of the IC instabilities, 
$t\gg \gamma^{-1}\left(\mathbf{k}\right)> |\omega^{-1}\left(\mathbf{k}\right)|$, electric field 
(\ref{8}) becomes the random function of the initial phase $\theta
\left(\mathbf{k}\right)$. The motion of ions and electrons in this field has a form 
of the random scattering of particles by the turbulent electric field and mimics to the 
thermal motion of particles. 

The average effect of the mesoscale  
inhomogeneity of the microscale IC turbulence on the mesoscale evolution of the ion 
and electron distribution functions of the poloidal sheared flow  was considered in Ref.
\cite{Mikhailenko3}. The central point in this theory is the transformation of the velocity
$\mathbf{v}_{i}$ and coordinates $X_{i}$ and $Y_{i}$  to the new microturbulence-associated 
velocity field $\tilde{\mathbf{v}}_{i}$ and  coordinates $\tilde{X}_{i}$, $\tilde{Y}_{i}$
determined by the relations\cite{Mikhailenko3}
\begin{eqnarray}
&\displaystyle
\tilde{\mathbf{v}}_{i}=\mathbf{v}_{i}-\tilde{\mathbf{V}}_{i}\left( X_{i}, Y_{i}, T, 
\varepsilon\right), 
\label{13}
\end{eqnarray}
\begin{eqnarray}
&\displaystyle
\tilde{X}_{i}=X_{i}-\int\limits^{t}_{t_{0}}\tilde{V}_{ix}\left( X_{i}, Y_{i}, T_{1}, 
\varepsilon\right)dT_{1}, 
\label{14}
\end{eqnarray}
\begin{eqnarray}
&\displaystyle
\tilde{Y}_{i}=Y_{i}-\int\limits^{T}_{T_{0}}\tilde{V}_{iy}\left( X_{i}, Y_{i}, T_{1}, \varepsilon
\right)dT_{1}, 
\label{15}
\end{eqnarray}
or by their inverse,
\begin{eqnarray}
&\displaystyle
\mathbf{v}_{i}=\tilde{\mathbf{v}}_{i}+\tilde{\mathbf{U}}_{i}
\left(\tilde{X}_{i}, \tilde{Y}_{i}, T, \varepsilon\right), 
\label{16}
\end{eqnarray}
\begin{eqnarray}
&\displaystyle
X_{i}=\tilde{X}_{i}+\tilde{R}_{ix}\left(\tilde{X}_{i}, \tilde{Y}_{i}, T, \varepsilon\right)=
\nonumber 
\\ 
&\displaystyle
=\tilde{X}_{i}+\int\limits^{T}_{T_{0}}
\tilde{U}_{ix}\left(\tilde{X}_{i}, \tilde{Y}_{i}, T_{1}, \varepsilon\right)dT_{1}, 
\label{17}
\end{eqnarray}
\begin{eqnarray}
&\displaystyle
Y_{i}=Y-V'_{0}XT 
\nonumber 
\\ 
&\displaystyle
=\tilde{Y}_{i}+\int\limits^{T}_{T_{0}}
\tilde{U}_{iy}\left(\tilde{X}_{i}, \tilde{Y}_{i}, T_{1}, \varepsilon\right)dT_{1}, 
\label{18}
\end{eqnarray}
where $V'_{0}$ is the velocity shear of the poloidal flow velocity $V_{0}\left(X\right)=V'_{0}X$
directed along coordinate $Y$. The velocity $\tilde{\mathbf{V}}_{i}\left( X_{i}, Y_{i}, T, 
\varepsilon\right)$ is determined by the Euler equation  
\begin{eqnarray}
&\displaystyle 
\varepsilon\frac{\partial \tilde{\mathbf{V}}_{i}}{\partial T}+\varepsilon\tilde{V}_{ix}
\frac{\partial\tilde{\mathbf{V}}_{i}\left( X_{i}, Y_{i}, T, \varepsilon\right)}{\partial {X}_{i}}
\nonumber 
\\ 
&\displaystyle
=\frac{e_{i}}{ m_{i}}\left(\tilde{\mathbf{E}}_{i}\left( X_{i}, Y_{i}, T, \varepsilon\right)
+\frac{1}{c}\left[\tilde{\mathbf{V}}_{i}\times\mathbf{B}_{0}\right]\right)
\label{19}
\end{eqnarray}
as the velocity of an ion in the electric field $\tilde{\mathbf{E}}_{i}\left(X_{i}, Y_{i}, T, 
\varepsilon\right)$ of the IC parametric turbulence, 
where variables $X_{i}$ and $Y_{i}$ are determined in the frame of references 
which moves with the velocity of an ion in the FW field in the poloidal 
sheared flow\cite{Mikhailenko3}. In variables $\left(\tilde{X}_{i}, \tilde{Y}_{i}, T\right)$,
the convective nonlinear part $\tilde{V}_{ix}\frac{\partial}{\partial {X}_{i}}$ of the operator 
$\frac{\partial}{\partial T}+\tilde{V}_{ix}\frac{\partial }{\partial 
{X}_{i}}$ of Eq. (\ref{19}) vanishes and this operator is transformed to the linear one, $
\frac{\partial}{\partial T}$. Then, Eq. (\ref{19}) becomes the ordinary differential 
equation 
\begin{eqnarray}
&\displaystyle 
\varepsilon\frac{d}{dT}\tilde{\mathbf{U}}_{i}\left(\tilde{X}_{i}, \tilde{Y}_{i}, T, 
\varepsilon\right)
\nonumber\\ 
&\displaystyle
=\frac{e_{i}}{m_{i}}\left(\tilde{\mathbf{E}}_{i}\left(\tilde{X}_{i}+
\tilde{R}_{ix}\left(\tilde{X}_{i}, \tilde{Y}_{i}, T, \varepsilon\right), \tilde{Y}_{i}, T, 
\varepsilon\right)\right.
\nonumber 
\\ 
&\displaystyle
\left.+\frac{1}{c}\left[\tilde{\mathbf{U}}_{i}\left(\tilde{X}_{i}, \tilde{Y}_{i}, T, 
\varepsilon\right)\times\mathbf{B}_{0}\right]\right).
\label{20}
\end{eqnarray}
for $\tilde{\mathbf{U}}_{i}\left(\tilde{X}_{i}, \tilde{Y}_{i}, T, \varepsilon\right)
= \tilde{\mathbf{V}}_{i}\left( X_{i}, Y_{i}, T, \varepsilon\right)$. The solution to Eq. (\ref{20})
may be easily derived\cite{Mikhailenko3} for the case of the small displacement, 
$\left|\tilde{R}_{ix}\right| \ll L_{\tilde{E}}$, of an ion in the inhomogeneous electric 
field $\tilde{\mathbf{E}}_{i}$. With the approximation for the amplitude $\tilde{\mathbf{E}}_{i}
\left(\mathbf{k}, X_{i}, n\right)$ of the $n$-th harmonic of the IC electric field 
in Eq. (\ref{20})
\begin{eqnarray}
&\displaystyle 
\tilde{\mathbf{E}}_{i}\left(\mathbf{k}, \tilde{X}_{i}+\tilde{R}_{ix}\left(\tilde{X}_{i}, 
\tilde{Y}_{i}, T, \varepsilon\right), n\right)\approx 
\tilde{\mathbf{E}}_{i}\left(\mathbf{k}, \tilde{X}_{i}, n\right),
\label{21}
\end{eqnarray}
the solution to Eq. (\ref{20}) with the initial value 
$\tilde{\mathbf{U}}_{i}\left(\tilde{X}_{i}, \tilde{Y}_{i}, T_{0}=0\right)=0$ is 
\begin{eqnarray}
&\displaystyle 
\tilde{U}_{ix}\left(\tilde{X}_{i}, \tilde{Y}_{i}, T, \varepsilon\right) 
= \frac{e_{i}}{\varepsilon m_{i}}\sum \limits_{n}\int\limits^{T}_{0}dT_{1}
\nonumber 
\\ &\displaystyle
\times\left[\tilde{E}_{ix}
\left(\tilde{X}_{i}, \tilde{Y}_{i}, n, T, \varepsilon\right)\cos \frac{1}{\varepsilon}\omega_{ci}
\left(T-T_{1}\right)\right. 
\nonumber 
\\ &\displaystyle
\left.+ \tilde{E}_{iy}
\left(\tilde{X}_{i}, \tilde{Y}_{i}, n, T, \varepsilon\right)\sin \frac{1}{\varepsilon}\omega_{ci}
\left(T-T_{1}\right)\right], 
\label{22}
\end{eqnarray}
\begin{eqnarray}
&\displaystyle 
\tilde{U}_{iy}\left(\tilde{X}_{i}, \tilde{Y}_{i}, T, \varepsilon\right) 
= \frac{e_{i}}{\varepsilon m_{i}}\sum \limits_{n}\int\limits^{T}_{0}dT_{1}
\nonumber 
\\ &\displaystyle
\times\left[-\tilde{E}_{ix}
\left(\tilde{X}_{i}, \tilde{Y}_{i}, n, T, \varepsilon\right)\sin \frac{1}{\varepsilon}\omega_{ci}
\left(T-T_{1}\right)\right. 
\nonumber 
\\ &\displaystyle
\left.+ \tilde{E}_{iy}
\left(\tilde{X}_{i}, \tilde{Y}_{i}, n, T, \varepsilon\right)\cos \frac{1}{\varepsilon}\omega_{ci}
\left(T-T_{1}\right)\right], 
\label{23}
\end{eqnarray}
With variables $\tilde{\mathbf{v}}_{i}$, $\tilde{X}_{i}, \tilde{Y}_{i}, Z_{i}, T, 
\varepsilon$, where $Z_{i}=\varepsilon z$, the Vlasov equation for the distribution function 
$F_{i}\left(\tilde{\mathbf{v}}_{i}, \tilde{X}_{i}, \tilde{Y}_{i}, Z_{i}, T, \varepsilon\right)$ 
of ions in the sheared poloidal flow for time $T\gg \tau_{corr} \sim \gamma^{-1}$  becomes
\begin{widetext}
\begin{eqnarray}
&\displaystyle 
\varepsilon\frac{\partial F_{i}}{\partial T} + \varepsilon\tilde{v}_{ix}
\frac{\partial F_{i}}{\partial \tilde{X}_{i}} 
+\varepsilon\left(\tilde{v}_{iy}-V'_{0}T\tilde{v}_{ix}\right)\frac{\partial F_{i}}
{\partial \tilde{Y}_{i}} +\varepsilon v_{iz}\frac{\partial F_{i}}{\partial Z_{i}}- 
\varepsilon\tilde{v}_{ix}\int\limits_{0}^{T}\frac{\partial}
{\partial X_{i}}\tilde{V}_{ix}\left(X_{i}, Y_{i}, T_{1}, \varepsilon\right)dT_{1}
\frac{\partial F_{i}}{\partial \tilde{X}_{i}} 
\nonumber 
\\ &\displaystyle
-\varepsilon\tilde{v}_{ix}\int\limits_{0}^{T}\frac{\partial}{\partial X_{i}}\tilde{V}_{iy}
\left(X_{i}, Y_{i}, T_{1}, \varepsilon \right)dT_{1}
\frac{\partial F_{i}}{\partial \tilde{Y}_{i}} 
\nonumber 
\\ &\displaystyle
- \varepsilon\tilde{U}_{ix}\left(\tilde{X}_{i}, 
\tilde{Y}_{i}, T, \varepsilon\right)\int\limits_{0}^{T}\frac{\partial}
{\partial X_{i}}\tilde{V}_{ix}\left(X_{i}, Y_{i}, T_{1}, \varepsilon\right)dT_{1}
\frac{\partial F_{i}}{\partial \tilde{X}_{i}}
\nonumber 
\\ &\displaystyle 
-\varepsilon\tilde{U}_{ix}\left(\tilde{X}_{i}, \tilde{Y}_{i}, T, \varepsilon\right)\left(V'_{0}T
+\int\limits_{0}^{T}\frac{\partial}{\partial X_{i}}\tilde{V}_{iy}
\left(X_{i}, Y_{i}, T_{1}, \varepsilon\right)dT_{1}\right)\frac{\partial F_{i}}
{\partial\tilde{Y}_{i}}
\nonumber     
\\ &\displaystyle
+\omega_{ci}\tilde{v}_{iy}\frac{\partial F_{i}}{\partial \tilde{v}_{ix}} -\omega_{ci}
\tilde{v}_{ix}\frac{\partial F_{i}}{\partial \tilde{v}_{iy}}
\nonumber 
\\ &\displaystyle
-\varepsilon\frac{e_{i}}{m_{i}}\left(\frac{\partial}{\partial \tilde{X}_{i}}
\varphi_{i}\left(\tilde{\mathbf{r}}_{i}+\tilde{\mathbf{R}}_{i}, \tilde{X}_{i}, 
\tilde{Y}_{i}, t \right)
- V'_{0}T \frac{\partial }{\partial \tilde{Y}_{i}}\varphi_{i}\left(\tilde{\mathbf{r}}_{i}+
\tilde{\mathbf{R}}_{i}, \tilde{X}_{i}, \tilde{Y}_{i}, t \right)\right)\frac{\partial F_{i}}{\partial 
\tilde{v}_{ix}}
\nonumber 
\\ &\displaystyle
-\varepsilon\frac{e_{i}}{m_{i}}\frac{\partial }{\partial \tilde{Y}_{i}}\varphi_{i}
\left(\tilde{\mathbf{r}}_{i}+
\tilde{\mathbf{R}}_{i}, \tilde{X}_{i}, \tilde{Y}_{i}, T\right)
\frac{\partial F_{i}}{\partial \tilde{v}_{iy}}
-\varepsilon\frac{e_{i}}{m_{i}}\frac{\partial }{\partial Z_{i}}\varphi_{i}
\left(\tilde{\mathbf{r}}_{i}+\tilde{\mathbf{R}}_{i}, \tilde{X}_{i}, \tilde{Y}_{i}, t\right)
\frac{\partial F_{i}}{\partial v_{iz}}=0.
\label{24}
\end{eqnarray}
\end{widetext}
The electrostatic potential $\varphi_{i}$ in Eq. (\ref{24}) depends on the micro- and mesoscales 
and can be expressed in the form
\begin{eqnarray}
&\displaystyle
\varphi_{i}\left(\tilde{\mathbf{r}}_{i}+\tilde{\mathbf{R}}_{i}, \tilde{X}_{i}, \tilde{Y}_{i}, t
\right)= \tilde{\varphi}_{i}
\left(\tilde{\mathbf{r}}_{i}
+\tilde{\mathbf{R}}_{i}, \tilde{X}_{i}, t\right)
\nonumber 
\\ &\displaystyle
+\Phi_{i}\left(\tilde{X}
_{i}, \tilde{Y}_{i}, Z_{i}, T\right),
\label{25}
\end{eqnarray}
where $\tilde{\varphi}_{i}$ is the electrostatic potential of the microscale turbulence, \\
\begin{eqnarray}
&\displaystyle 
\tilde{\mathbf{E}}_{i}\left( \tilde{X}_{i}, \tilde{Y}_{i}, T, \varepsilon\right)
=-\nabla_{\mathbf{r}_{i}}\varphi_{i}\left(\tilde{\mathbf{r}}_{i}+\tilde{\mathbf{R}}_{i}, 
\tilde{X}_{i}, t\right),
\label{26}
\end{eqnarray} 
and $\Phi_{i}\left(\tilde{X}_{i}, 
\tilde{Y}_{i}, Z_{i}, T\right)$ is the potential which determines the electric field 
of the plasma response on the mesoscale convective flows, 
\begin{eqnarray}
&\displaystyle
\bar{\mathbf{E}}_{i}\left(\tilde{X}_{i}, 
\tilde{Y}_{i}, Z_{i}, T\right)=-\nabla\Phi_{i}\left(\tilde{X}_{i}, \tilde{Y}_{i}, Z_{i}, T\right). 
\label{27}
\end{eqnarray}
The Vlasov equation for the ion distribution function 
$\bar{F}_{i}\left(\tilde{\mathbf{v}}_{i}, \tilde{X}_{i}, \tilde{Y}_{i}, Z_{i}, 
T, \varepsilon\right)$, averaged over the microscale initial phases for a time 
$t\gg \tau _{corr}\sim \gamma^{-1}$, is
\begin{widetext}
\begin{eqnarray}
&\displaystyle 
\frac{\partial \bar{F}_{i}}{\partial T} +\tilde{v}_{ix}\frac{\partial \bar{F}_{i}}
{\partial \tilde{X}_{i}} +\left(\tilde{v}_{iy}-V'_{0}T\tilde{v}_{ix}\right)
\frac{\partial \bar{F}_{i}}{\partial \tilde{Y}_{i}}-\bar{U}_{ix}\left(\tilde{X}_{i}\right)
\frac{\partial \bar{F}_{i}}{\partial \tilde{X}_{i}} -\bar{U}_{iy}\left(\tilde{X}_{i}\right)
\frac{\partial \bar{F}_{i}}{\partial \tilde{Y}_{i}}
\nonumber 
\\ &\displaystyle
+v_{iz}\frac{\partial \bar{F}_{i}}{\partial Z_{i}}
+\frac{1}{\varepsilon}\omega_{ci}\tilde{v}_{iy}\frac{\partial \bar{F}_{i}}{\partial \tilde{v}_{ix}}
-\frac{1}{\varepsilon}\omega_{ci}\frac{\partial \bar{F}_{i}}{\partial \tilde{v}_{iy}}
\nonumber 
\\ &\displaystyle
-\frac{e_{i}}{m_{i}}\left(\frac{\partial \Phi_{i} \left(\tilde{X}_{i}, \tilde{Y}_{i}, 
Z_{i}, T\right)}
{\partial \tilde{X}_{i}}- V'_{0}T \frac{\partial \Phi_{i} \left(\tilde{X}_{i}, 
\tilde{Y}_{i}, Z_{i}, T\right)}{\partial \tilde{Y}_{i}}\right)\frac{\partial F_{i}}{\partial 
\tilde{v}_{ix}}
\nonumber 
\\ &\displaystyle
-\frac{e_{i}}{m_{i}}\frac{\partial \Phi_{i} \left(\tilde{X}_{i}, \tilde{Y}_{i}, Z_{i}, T\right)}
{\partial \tilde{Y}_{i} }\frac{\partial F_{i}}{\partial \tilde{v}_{iy}}
-\frac{e_{i}}{m_{i}}\frac{\partial \Phi_{i} \left(\tilde{X}_{i}, \tilde{Y}_{i}, Z_{i}, T\right)}
{\partial Z_{i}}\frac{\partial F_{i}}{\partial v_{iz}}=0,
\label{28}
\end{eqnarray}
where the velocities $\bar{U}_{ix}\left(\tilde{X}_{i}\right)$ and $\bar{U}_{iy}
\left(\tilde{X}_{i}\right)$ are
\begin{eqnarray}
&\displaystyle 
\bar{U}_{ix}\left(\tilde{X}_{i}\right)=\left\langle\tilde{U}_{ix}\left(\tilde{X}_{i}, 
\tilde{Y}_{i}, T, \varepsilon\right)
\int\limits_{0}^{T}\frac{\partial}
{\partial X_{i}}\tilde{V}_{ix}\left(X_{i}, Y_{i}, T_{1}, \varepsilon\right)dT_{1}
\right\rangle, 
\label{29}
\end{eqnarray}
\begin{eqnarray}
&\displaystyle 
\bar{U}_{iy}\left(\tilde{X}_{i}\right)=\left\langle\tilde{U}_{ix}\left(\tilde{X}_{i}, 
\tilde{Y}_{i}, T, \varepsilon\right)\int\limits_{0}^{T}\frac{\partial}
{\partial X_{i}}\tilde{V}_{iy}\left(X_{i}, Y_{i}, T_{1}, \varepsilon\right)dT_{1}
\right\rangle. 
\label{30}
\end{eqnarray}
\end{widetext}
The details of the calculation of $\bar{U}_{ix}\left(\tilde{X}_{i}\right)$ and $\bar{U}_{iy}
\left(\tilde{X}_{i}\right)$ for the arbitrary electric field $\tilde{\mathbf{E}}_{i}$ are 
given in Ref. \cite{Mikhailenko1}. For the electric field (\ref{8}), the velocities 
$\bar{U}_{ix}\left(\tilde{X}_{i}, T\right)$ and $\bar{U}_{iy}\left(\tilde{X}_{i}, T\right)$ 
are presented in the Appendix. 

The electrostatic potential $\Phi_{i}\left(\tilde{X}_{i}, 
\tilde{Y}_{i}, Z_{i}, T\right)$ of the plasma response on the mesoscale convective flows
is governed by the Poisson equation
\begin{eqnarray}
&\displaystyle 
\frac{\partial^{2} \Phi_{i}\left(\tilde{X}_{i}, \tilde{Y}_{i}, Z_{i}, T, \varepsilon\right)}
{\partial \tilde{X_{i}^{2}}}+ \frac{\partial^{2}\Phi_{i}\left(\tilde{X}_{i}, \tilde{Y}_{i}, 
Z_{i}, T, \varepsilon\right)}{\partial\tilde{Y_{i}^{2}}}
\nonumber 
\\ &\displaystyle
+\frac{\partial^{2} \Phi_{i}
\left(\tilde{X}_{i}, \tilde{Y}_{i}, Z_{i}, T, \varepsilon\right)}
{\partial Z_{i}^{2}}
\nonumber 
\\ &\displaystyle
=-4\pi\sum_{\alpha=i, e}e_{\alpha}n_{\alpha}\left(\tilde{X}_{\alpha}, \tilde{Y}_{\alpha}, 
Z_{\alpha}, T, \varepsilon\right).
\label{31}
\end{eqnarray}
in which  $n_{\alpha}\left(\tilde{X}_{\alpha}, \tilde{Y}_{\alpha}, 
Z_{\alpha}, T, \varepsilon\right)=\int d\tilde{\mathbf{v}}_{\alpha}\bar{f}_{\alpha}
\left(\tilde{\mathbf{v}}_{\alpha}, \tilde{X}_{\alpha}, \tilde{Y}_{\alpha}, Z_{\alpha}, T, 
\varepsilon \right)$ is the density perturbation, and $\bar{f}_{\alpha}
\left(\tilde{\mathbf{v}}_{\alpha}, \tilde{X}_{\alpha}, \tilde{Y}_{\alpha}, Z_{\alpha}, T, 
\varepsilon \right)=\bar{F}_{\alpha}\left(\tilde{v}_{\alpha\perp}, 
\phi, v_{z}, \xi_{\alpha}, \eta_{\alpha}, Z_{\alpha}, T, \varepsilon\right)-\bar{F}_{\alpha0}$ 
is the perturbation of the equilibrium distribution function $\bar{F}_{\alpha0}$ of the 
convected plasma species $\alpha$.

The Vlasov equation for the average electron distribution $\bar{F}_{e}
\left(\tilde{\mathbf{v}}_{e}, \tilde{X}_{e}, \tilde{Y}_{e}, Z_{e}, T\right)$, where 
$\tilde{X}_{e}, \tilde{Y}_{e}$ are determined by Eqs. (\ref{14}), (\ref{15}) with ion species 
subscript changed on the electron species subscript, has a form similar to 
Eq. (\ref{28}). The velocities $\bar{U}_{ex}\left(\tilde{X}_{e}\right)$ and $\bar{U}_{ey}
\left(\tilde{X}_{e}\right)$ are determined in this equation by Eqs. (\ref{29}), (\ref{30}) in 
which velocities $\tilde{U}_{ex}\left(\tilde{\mathbf{r}}_{e}, \tilde{X}_{e}, T\right)$ and 
$\tilde{U}_{ey}\left(\tilde{\mathbf{r}}_{e}, \tilde{X}_{e}, T\right)$ are determined by Eqs. 
(\ref{22}), (\ref{23}) with the turbulent electric field $\tilde{\mathbf{E}}_{e}
\left(\tilde{\mathbf{r}}_{e}, \tilde{X}_{e}, T\right)$, given by Eq. (\ref{10}).

The Vlasov equations (\ref{28}) for $\bar{F}_{i}\left(\tilde{\mathbf{v}}_{i}, \tilde{X}_{i}, 
\tilde{Y}_{i}, Z_{i}, T\right)$, and the similar equation for \\ $\bar{F}_{e}
\left(\tilde{\mathbf{v}}_{e}, \tilde{X}_{e}, \tilde{Y}_{e}, Z_{e}, T\right)$, 
and the Poisson equation (\ref{31}) compose the Vlasov-Poisson system, which governs the 
kinetic mesoscale evolution of a plasma under the average action of the spatially
inhomogeneous microturbulence. 
 
As a first step to solution of the system of Eqs. (\ref{28}), (\ref{31}), 
we find the characteristics of the operator 
\begin{eqnarray}
&\displaystyle 
\frac{D}{DT}=\frac{\partial }{\partial T} -\bar{U}_{ix}\left(\tilde{X}_{i}\right)
\frac{\partial }
{\partial \tilde{X}_{i}} -\bar{U}_{iy}\left(\tilde{X}_{i}\right)\frac{\partial}
{\partial \tilde{Y}_{i}},
\label{32}
\end{eqnarray}
which are determined by the system
\begin{eqnarray}
&\displaystyle 
dT=-\frac{d\tilde{X}_{i}}{\bar{U}_{ix}\left(\tilde{X}_{i}\right)}=-\frac{d\tilde{Y}_{i}}
{\bar{U}_{iy}\left(\tilde{X}_{i}\right)}.
\label{33}
\end{eqnarray}
For deriving the simplest solution to system (\ref{33}), which reveals the effects of 
the spatial inhomogeneity of the convective flow velocities, we use in Eq. (\ref{33}) the 
expansions
\begin{eqnarray}
&\displaystyle \bar{U}_{ix}\left(\tilde{X}_{i}\right)=\bar{U}^{(0)}_{ix}+\bar{U}'_{ix}
\left(\tilde{X}^{(0)}_{i}\right)\left(\tilde{X}_{i}-\tilde{X}^{(0)}_{i}\right), 
\label{34}
\end{eqnarray}
and 
\begin{eqnarray}
&\displaystyle \bar{U}_{iy}\left(\tilde{X}_{i}\right) =\bar{U}^{(0)}_{iy}+ \bar{U}'_{iy}
\left(\tilde{X}^{(0)}_{i}\right)\left(\tilde{X}_{i}-\tilde{X}^{(0)}_{i}\right)
\label{35}
\end{eqnarray}
at the vicinity of an arbitrary coordinate $\tilde{X}^{(0)}_{i}$, 
and consider the case of the uniform velocity compressing rate,  
$\bar{U}'_{ix}= const$,  and of the  uniform velocity shearing rate, $\bar{U}'_{iy}= const$. 
The solution to system (\ref{33}) for this case has a form\cite{Mikhailenko3}
\begin{eqnarray}
&\displaystyle 
\check{X}_{i}=\frac{1}{\bar{U}'_{ix}}\left[\left(\bar{U}^{(0)}_{ix}+\bar{U}'_{ix}
\left(\tilde{X}_{i}- \tilde{X}^{(0)}_{i}\right) 
\right)e^{\bar{U}'_{ix}T}\right.
\nonumber 
\\ &\displaystyle
\left.- \bar{U}^{(0)}_{ix} \right], 
\label{36}
\end{eqnarray}
and
\begin{eqnarray}
&\displaystyle 
\check{Y}_{i}=\tilde{Y}_{i}+\left(\bar{U}^{(0)}_{iy}-\bar{U}^{(0)}_{ix}\frac{\bar{U}'_{iy}}
{\bar{U}'_{ix}}\right)T
\nonumber 
\\ 
&\displaystyle
-\frac{\bar{U}'_{iy}}{\left(\bar{U}'_{ix}\right)^{2}}\left(\bar{U}^{(0)}_{ix}
+\bar{U}'_{ix}\left(\tilde{X}_{i}-\tilde{X}^{(0)}_{i}\right)\right),
\label{37}
\end{eqnarray}
where $\check{X}_{i}$ and $\check{Y}_{i}$ are the integrals of system (\ref{33}) with 
expansions (\ref{34}), (\ref{35}). Note, that at $T=0$, $\check{X}_{i}=\tilde{X}_{i}
-\tilde{X}^{(0)}_{i}$. In what follows, we put $X_{0}=0$ for simplicity.

With variables $\check{X}_{i}, \check{Y}_{i}, Z_{i}, T, v_{z}, \varepsilon$ and with 
$\tilde{v}_{i\perp}, \phi$, determined by relations $\tilde{v}_{ix}=\tilde{v}_{i\perp
}\cos \phi$ and $\tilde{v}_{iy}=\tilde{v}_{i\perp}\sin \phi$, the Vlasov 
equation (\ref{28}) obtains the form, which does not contain the explicit dependence on 
$\tilde{X}_{i}$,
\begin{widetext}
\begin{eqnarray}
&\displaystyle 
\frac{\partial }{\partial T}\bar{F}_{i}\left(\tilde{v}_{i\perp}, \phi, v_{z}, \check{X}_{i}, 
\check{Y}_{i}, Z_{i}, T, \varepsilon\right)+\tilde{v}_{i\perp}\cos \phi\,e^{\bar{U}_{ix}'T}
\frac{\partial \bar{F}_{i}}
{\partial \check{X}_{i}} 
\nonumber 
\\ &\displaystyle
+\left(\tilde{v}_{i\perp}\sin\phi-
\tilde{v}_{i\perp}\cos\phi \,\left(V'_{0}T+\frac{\bar{U}_{iy}'}{\bar{U}_{ix}'}
\right)\right)\frac{\partial \bar{F}_{i}}{\partial \check{Y}_{i}}
+v_{iz}\frac{\partial \bar{F}_{i}}{\partial Z_{i}}
-\omega_{ci}\frac{\partial \bar{F}_{i}}{\varepsilon\partial \phi}
\nonumber 
\\ &\displaystyle
-\frac{e_{i}}{m_{i}}\left(e^{\bar{U}_{ix}'T}\frac{\partial \Phi_{i} \left(\check{X}_{i}, 
\check{Y}_{i}, 
Z_{i}, T\right)}{\partial \check{X}_{i}}- \left(V'_{0}T+\frac{\bar{U}'_{iy}}{\bar{U}'_{ix}}\right) 
\frac{\partial \Phi_{i} \left(\check{X}_{i}, 
\check{Y}_{i}, Z_{i}, T\right)}{\partial \check{Y}_{i}}\right)\frac{\partial \bar{F}_{i}}
{\partial\check{v}_{ix}}
\nonumber 
\\ &\displaystyle
-\frac{e_{i}}{m_{i}}\frac{\partial \Phi_{i} \left(\check{X}_{i}, \check{Y}_{i}, Z_{i}, T\right)}
{\partial \check{Y}_{i} }\frac{\partial \bar{F}_{i}}{\partial \tilde{v}_{iy}}
-\frac{e_{i}}{m_{i}}\frac{\partial \Phi_{i} \left(\check{X}_{i}, \check{Y}_{i}, Z_{i}, T\right)}
{\partial Z_{i}}\frac{\partial \bar{F}_{i}}{\partial v_{iz}}=0.
\label{38}
\end{eqnarray}
\end{widetext}
In Eq. (\ref{38}), the effects the spatial inhomogeneity of the sheared and convected flows 
is transformed to the time domain. These effects are presented by the linearly growing with time 
$V'_{0}T$ coefficient originated from the basic poloidal sheared flow, and of the exponentially 
growing with time coefficient $e^{\bar{U}_{ix}'T}$ originated from the compressed convective flow. 
These coefficients 
reveal the effects of the continuous distortion of the perturbations in the sheared and 
compressed flows \cite{Mikhailenko2,Mikhailenko4,Mikhailenko5}.  It was found in Ref. 
\cite{Mikhailenko3} that the compressing rate $\bar{U}'_{ix}$ of the radial velocity of the 
compressed flow and the shearing rate $V'_{0}$ of the poloidal sheared flow velocity are 
commensurable for the tokamak edge condition, and both are much less 
than the ion cyclotron frequency $\omega_{ci}$. Equation (\ref{38}) displays that the 
exponentially growing with time effect of the compressed flow is the dominant factor in 
the mesoscale temporal evolution at time $T> \left(\bar{U}'_{ix}\right)^{-1}$ of the plasma 
with a radially inhomogeneous turbulence. Therefore the small parameter $\varepsilon$ 
in Eq. (\ref{38}), which determines a ratio of the micro- to meso-  scales, is naturally to 
define as equal to $\varepsilon=\frac{\bar{U}'_{ix}}{\omega_{ci}}$.

The next substantial simplification of Eq. (\ref{38}) arises with transformation of the part 
of Eq. (\ref{38}), which does not contain the electrostatic potential $\Phi_{i}\left(\check{X}_{i}, 
\check{Y}_{i}, T\right)$, by employing the characteristic equations
\begin{eqnarray}
&\displaystyle 
dT=-\frac{\varepsilon d\phi}{\omega_{ci}}=\frac{dZ_{i}}{v_{z}}
=\frac{d\check{X}_{i}}{\tilde{v}_{i\perp}\cos \phi\,e^{\bar{U}_{ix}'T}}
\nonumber 
\\ &\displaystyle
=\frac{d\check{Y}_{i}}{\tilde{v}_{i\perp}\sin\phi-
\tilde{v}_{i\perp}\cos\phi \,\left(V'_{0}T+\frac{\bar{U}_{iy}'}{\bar{U}_{ix}'}\right)}.
\label{39}
\end{eqnarray}
The solutions to Eqs. (\ref{39}) are given by the relations
\begin{eqnarray}
&\displaystyle 
\check{X}_{i}=\xi_{i}-\frac{\tilde{v}_{i\perp}\varepsilon}{\omega_{ci}}e^{\bar{U}_{ix}'T}\sin
\left(\phi_{1}-
\frac{1}{\varepsilon}\omega_{ci}T\right)
\nonumber 
\\ &\displaystyle
+O\left(\frac{\varepsilon\bar{U}'_{ix}}
{\omega_{ci}}\ll 
1\right),
\label{40}
\end{eqnarray}
\begin{eqnarray}
&\displaystyle 
\check{Y}_{i}=\eta_{i}+\frac{\tilde{v}_{i\perp}\varepsilon}{\omega_{ci}}\cos\left(\phi_{1}-\frac{1}
{\varepsilon}\omega_{ci}T\right)
\nonumber 
\\ &\displaystyle
+\frac{\tilde{v}_{i\perp}\varepsilon}{\omega_{ci}}\left(V'_{0}T+\frac{\bar{U}_{iy}'}{\bar{U}
_{ix}'}\right)\sin\left(\phi_{1}-\frac{1}{\varepsilon}\omega_{ci}T\right)
\nonumber 
\\ &\displaystyle
+O\left(\varepsilon
\frac{V'_{0}}{\omega_{ci}}\ll 1\right),
\label{41}
\end{eqnarray}
\begin{eqnarray}
&\displaystyle 
\phi=\phi_{1}-\frac{1}{\varepsilon}\omega_{ci}T,
\label{42}
\end{eqnarray}
\begin{eqnarray}
&\displaystyle 
Z_{i}=Z_{i1}+v_{z}T.
\label{43}
\end{eqnarray}
The integrals $\xi_{i}$ and $\eta_{i}$ in Eqs. (\ref{40}) and (\ref{41}) are the guiding 
center coordinates in the compressed-sheared convective flow. In coordinates 
$\xi_{i},\,\eta_{i}, \phi_{1}, Z_{i1}$, Eq. (\ref{38}) has a simple form
\begin{widetext}
\begin{eqnarray}
&\displaystyle 
\frac{\partial}{\partial T}\bar{F}_{i}\left(\tilde{v}_{i\perp}, \phi_{1}, v_{z},
\xi_{i}, \eta_{i}, Z_{i}, T, \varepsilon\right)+\frac{e_{i}}{m_{i}}\frac{\omega_{ci}}
{\varepsilon\tilde{v}_{i\perp}}
\left(\frac{\partial \Phi_{i}}{\partial \phi_{1}}\frac{\partial \bar{F}_{i}}{\partial 
\tilde{v}_{i\perp}}-\frac{\partial \Phi_{i}}{\partial \tilde{v}_{i\perp}}\frac{\partial \bar{F}_{i}}
{\partial \phi_{1}}\right)
\nonumber 
\\ &\displaystyle
+\frac{e_{i}\varepsilon}{m_{i}\omega_{ci}}e^{\bar{U}'_{ix}T}\left(\frac{\partial \Phi_{i}}{\partial 
\xi_{i}}\frac{\partial \bar{F}_{i}}{\partial 
\eta_{i}}-\frac{\partial \Phi_{i}}{\partial \eta_{i}}\frac{\partial \bar{F}_{i}}{\partial \xi_{i}}
\right)-\frac{e_{i}}{m_{i}}\frac{\partial \Phi_{i}}{\partial Z_{i1}}\frac{\partial \bar{F}_{i}}
{\partial v_{z}}=0.
\label{44}
\end{eqnarray}
\end{widetext}
The solution to Eq. (\ref{44}) for the ion distribution function $\bar{F}_{i}$ we derive 
in the form $\bar{F}_{i}\left(\tilde{v}_{i\perp}, 
\phi, v_{z}, \xi_{i}, \eta_{i}, Z_{i1}, T, \varepsilon\right)=\bar{F}_{i0}
+\bar{f}_{i}\left(\tilde{v}_{i\perp}, \phi, v_{z}, \xi_{i}, \eta_{i}, Z_{i1}, T,
\varepsilon\right)$, where $\bar{F}_{i0}$ is the ion distribution function $\bar{F}_{i0}$ of the 
unperturbed ion convective flow, and $\bar{f}_{i}$ is the perturbation of $\bar{F}_{i0}$ 
caused by the ions respond on the plasma convective flows. The equation for $\bar{F}_{i0}$ 
follows from Eq. (\ref{44}), in which potential $\Phi_{i}$ of the 
electrostatic response of plasma on the mesoscale convective flows is excluded. This equation,
\begin{eqnarray}
&\displaystyle 
\partial \bar{F}_{i0}/\partial T=0, 
\label{45}
\end{eqnarray}
reveals that with the guiding center coordinates $\xi_{i}$, $\eta_{i}$, the unperturbed  
ion distribution function $\bar{F}_{i0}$ of the compressed-sheared ion flow is 
stationary. The solution to Eq. (\ref{45}) for the inhomogeneous ion component along coordinate 
$\tilde{X}_{i}$ is an arbitrary function $\bar{F}_{i0}=\bar{F}_{i0}\left(\tilde{\mathbf{v}}_{i}, 
\xi_{i}\right)$. With the initial Maxwellian distribution 
\begin{eqnarray}
&\displaystyle 
\bar{F}_{i0}\left(\tilde{\mathbf{v}}_{i}, \tilde{X}_{i}\right)=
\frac{n_{i0}\left(\tilde{X}_{i}\right)}{\left(2\pi v^{2}_{Ti}\left(\tilde{X}_{i}\right)
\right)^{3/2}}e^{-\frac{\tilde{v}^{2}_{i}}{2v^{2}_{Ti}\left(\tilde{X}_{i}\right)}},
\label{46}
\end{eqnarray}
given for time $T=0$, at which $\xi_{i}\approx \check{X}_{i}=\tilde{X}_{i}$ (it follows from 
Eq. (\ref{36}) and from the estimate $\tilde{v}_{i}/\omega_{ci}\sim v_{Ti}/\omega_{ci}
\ll \left(L_{E}, L_{n_{i}}\right)$) the solution for $\bar{F}_{i0}$ will have a form (\ref{46}), 
in which the ion equilibrium density and the ion thermal velocity are equal to 
$n_{i0}\left(\xi_{i}\right)$ and $v_{Ti}\left(\xi_{i}\right)$ respectively. Note, that 
with variable $\tilde{X}_{i}$ the ion density $n_{i0}$ is the time dependent,
\begin{eqnarray}
&\displaystyle 
n_{i0}\left(\xi_{i}\right)=n_{i0}\left(
\frac{1}{\bar{U}'_{ix}}\left[\left(\bar{U}^{(0)}_{ix}+\bar{U}'_{ix}
\tilde{X}_{i}\right)e^{\bar{U}'_{ix}T}\right.\right.
\nonumber 
\\ &\displaystyle
\left.\left.- \bar{U}^{(0)}_{ix} \right]\right).
\label{47}
\end{eqnarray}
The same dependences on $\tilde{X}_{i}$ and $T$ has  the ion thermal velocity
$v_{Ti}\left(\xi_{i}\right)$ in Eq. (\ref{46}). 
Equation (\ref{46}) reveals, that the time dependent inhomogeneous ion density of the 
convective flow, as it is with coordinate $\tilde{X}_{i}$, becomes the steady spatially 
inhomogeneous ion density distribution along characteristic $\xi_{i}$ 
at any time $T$. 

\section{The compressed-sheared modes approach to the Stability theory of the 
mesoscale convective flows}\label{sec4} 

In this section, we consider the microscale respond of the ions and electrons, determined by the 
functions $\bar{f}_{i}$ and $\bar{f}_{e}$ on the generation of the  mesoscale convective flows.

It follows from Eq. (\ref{44}), that the Vlasov equation for the perturbation \\ $\bar{f}_{i}
\left(\tilde{v}_{i\perp}, \phi, v_{iz}, \xi_{i}, \eta_{i}, Z_{i1}, T,
\varepsilon\right)$ of $\bar{F}_{i0}\left(\tilde{\mathbf{v}}_{i}, 
\xi_{i}\right)$
has a simple form in guiding center coordinates $\xi_{i}$ and $\eta_{i}$, 
\begin{eqnarray}
&\displaystyle 
\frac{\partial}{\partial T}\bar{f}_{i}\left(\tilde{v}_{i\perp}, \phi_{1}, v_{iz},\xi_{i}, 
\eta_{i}, Z_{i1}, T, \varepsilon\right)
=-\frac{e_{i}}{m_{i}}\frac{\omega_{ci}}{\varepsilon\tilde{v}_{i\perp}}
\frac{\partial \Phi_{i}}{\partial \phi_{1}}\frac{\partial \bar{F}_{i0}}{\partial 
\tilde{v}_{i\perp}}
\nonumber 
\\ &\displaystyle
+\frac{e_{i}\varepsilon}{m_{i}\omega_{ci}}e^{\bar{U}'_{ix}T}\frac{\partial \Phi_{i}}
{\partial \eta_{i}}\frac{\partial \bar{F}_{i0}}{\partial \xi_{i}}+\frac{e_{i}}{m_{i}}
\frac{\partial \Phi_{i}}{\partial Z_{i1}}\frac{\partial \bar{F}_{i0}}{\partial v_{iz}}.
\label{48}
\end{eqnarray}
The Vlasov equation (\ref{48}) for $\bar{f}_{i}$ with given unperturbed  
ion distribution function $\bar{F}_{i0}$, the equation for the perturbation $\bar{f}_{e}
\left(\tilde{v}_{e\perp}, \phi_{1}, v_{z}, \xi_{e}, \eta_{e}, Z_{e1}, T, 
\varepsilon\right)$ of the electron distribution similar to Eq. (\ref{48}), 
and the Poisson equation (\ref{35}) for the potential $\Phi_{i}\left(\check{X}
_{i}, \check{Y}_{i}, Z_{i1}, T, \varepsilon\right)$ in coordinates $\check{X}_{i},\, 
\check{Y}_{i}$ 
\begin{eqnarray}
&\displaystyle 
e^{2\bar{U}_{ix}'T}\frac{\partial^{2} \Phi_{i}}{\partial \check{X}^{2}_{i}}
-2e^{\bar{U}_{ix}'T}\left(V'_{0}T+\frac{\bar{U}_{iy}'}{\bar{U}_{ix}'}\right)\frac{\partial^{2} 
\Phi_{i}}{\partial 
\check{X}_{i}\partial \check{Y}_{i}}
\nonumber 
\\ &\displaystyle
+\left(1+\left(V'_{0}T+\frac{\bar{U}_{iy}'}{\bar{U}_{ix}'}\right)^{2}\right)
\frac{\partial^{2}\Phi_{i}}
{\partial \check{Y}^{2}_{i}}+\frac{\partial^{2} \Phi_{i}}{\partial Z^{2}_{i1}}
\nonumber 
\\ &\displaystyle
=-4\pi\left(e_{i}n_{i}\left(\check{X}_{i}, \check{Y}_{i}, Z_{i1}, T; \varepsilon\right)\right.
\nonumber 
\\ &\displaystyle
\left.-|e|n_{e}\left(\check{X}_{e}, \check{Y}_{e}, Z_{e1}, T; \varepsilon\right)\right).
\label{49}
\end{eqnarray}
compose the system of equations for the investigations of the stability of the mesoscale  
convective flows. In this section, we consider the stability of the compressed-sheared 
flows against the development of the low frequency microscale instabilities. For that goal, we 
transform Eqs. (\ref{48}), (\ref{49}) to the microscale time $t=\frac{T}{\varepsilon}$ , the 
microscale spatial coordinates $x_{i}=\frac{1}{\varepsilon}X_{i}$,  $y_{i}=\frac{1}{\varepsilon}Y_{i}$ 
and to microscale coordinates guiding center coordinates $\check{\xi}_{i}=\frac{\xi_{i}}{\varepsilon}
$, $\check{\eta}_{i}=\frac{\eta_{i}}{\varepsilon}$. With time $t$ and microscale coordinates 
$\check{\xi}_{i}$, $\check{\eta}_{i}$, the solution to Eq. (\ref{48}) with known distribution 
$\bar{F}_{i0}$ is 
\begin{eqnarray}
&\displaystyle 
\bar{f}_{i}=\frac{e_{i}}{m_{i}}\int\limits^{t}_{0}dt_{1}\left(\frac{
e^{\bar{U}'_{ix}t}}{\omega_{ci}}\frac{\partial \Phi_{i}}{\partial \check{\eta}_{i}}
\frac{\partial \bar{F}_{i0}}{\partial \check{\xi}_{i}}\right.
\nonumber 
\\ &\displaystyle
\left.-\frac{\omega_{ci}}{\tilde{v}_{i\perp}}
\frac{\partial \Phi_{i}}{\partial \phi_{1}}\frac{\partial \bar{F}_{i0}}{\partial 
\tilde{v}_{i\perp}}+\frac{\partial \Phi_{i}}{\partial Z_{i1}}
\frac{\partial \bar{F}_{i0}}{\partial v_{z}}\right),
\label{50}
\end{eqnarray}
where the prime in $\bar{U}'_{ix}$, $\bar{U}'_{iy}$ and $V'_{0}$ denotes in this 
section the derivatives of $\bar{U}_{ix}$, $\bar{U}_{iy}$ and $V_{0}$ with respect 
to the microscale coordinate 
$\tilde{x}_{i}=\frac{\tilde{X}_{i}}{\varepsilon}$.  Equation (\ref{50}), as 
well as Eq. (\ref{44}), do not contain the spatial inhomogeneity originated 
from the inhomogeneity of the convective flows velocities.
Therefore, by the Fourier transforming the potential $\Phi_{i}\left(\check{x}_{i}, 
\check{y}_{i}, z_{i1}, t\right)$ over the microscale spatial coordinates $\check{x}_{i}, 
\check{y}_{i}$, 
\begin{eqnarray}
&\displaystyle 
\Phi_{i}\left(\check{x}_{i}, \check{y}_{i}, z_{i1}, t, \right)=\frac{1}{\left(2\pi\right)^{3}}\int 
dk_{\check{x}_{i}}dk_{\check{y}_{i}}dk_{z}
\nonumber 
\\ &\displaystyle
\times\Phi_{i}\left(k_{\check{x}_{i}}, k_{\check{y}_{i}}, k_{z}, t\right)e^{i
\left(k_{\check{x}_{i}}\check{x}_{i}+ k_{\check{y}_{i}}\check{y}_{i}+k_{z}z_{i}\right)}
\label{51}
\end{eqnarray}
we will derive from Eqs. (\ref{48}) and (\ref{49}) the equation for {\textit{the separate spatial 
Fourier mode $\Phi_{i}\left(k_{\check{x}_{i}}, k_{\check{y}_{i}}, k_{z}, t\right)$}} of the 
microscale plasma response on the mesoscale compressed-sheared convective flows. With coordinates 
$\check{\xi}_{i}$, $\check{\eta}_{i}$, used in Eq. (\ref{50}), potential $\Phi_{i}\left(\check{\xi}
_{i}, \check{\eta}_{i}, z_{i1}, t\right)$ is determined by the relation
\begin{eqnarray}
&\displaystyle 
\Phi_{i}\left(\check{x}_{i}, \check{\eta}_{i}, z_{i1}, t\right)=
\frac{1}{\left(2\pi\right)^{3}}\int dk_{\check{x}_{i}}
dk_{\check{y}_{i}}dk_{Z}
\nonumber 
\\ &\displaystyle
\times \Phi_{i}\left(k_{\check{x}_{i}}, k_{\check{y}_{i}}, k_{z}, t\right)e^{i\left(k_{\check{x}_{i}}
\check{\xi}_{i}+k_{\check{y}_{i}}\check{\eta}_{i}+k_{z}z_{i1}\right)}
\nonumber 
\\ &\displaystyle
\times\exp\left(-i\frac{k_{i\bot}\left(t\right)\tilde{v}_{i\perp}}{\omega_{ci}}
\sin\left(\phi-\omega_{ci}t-\delta\left(t\right)\right)\right)
\nonumber 
\\ &\displaystyle
=\int dk_{\check{x}_{i}}dk_{\check{y}_{i}}dk_{z}
\Phi_{i}\left(k_{\check{x}_{i}}, k_{\check{y}_{i}}, k_{z}, t\right)
\nonumber 
\\ &\displaystyle
\times e^{i\left(k_{\check{x}_{i}}\check{\xi}_{i}+ k_{\check{y}_{i}}
\check{\eta}_{i}+k_{z}z_{i1}\right)}
\nonumber 
\\ &\displaystyle
\times\sum\limits_{n=-\infty}^{\infty}J_{n}\left(\frac{k_{i\bot}\left(t\right)
\tilde{v}_{i\perp}}{\omega_{ci}}\right)
\nonumber 
\\ &\displaystyle
\times
e^{-in\left(\phi_{1}-
\omega_{ci}t-\delta\left(t\right)\right)},
\label{52}
\end{eqnarray}
in which $J_{n}$ is the Bessel function of the first kind of the order $n$ and 
the wave number component $k_{i\bot} \left(t\right)$ across the magnetic field grows 
with time due to the distorting of the wave structure by the compressed and sheared flows,
\begin{eqnarray}
&\displaystyle 
k^{2}_{i\bot}\left(t\right)=\left(k_{\check{x}_{i}}e^{\bar{U}'_{ix}t}-k_{\check{y}_{i}}
\left(V'_{0}t+\frac{\bar{U}_{iy}'}{\bar{U}_{ix}'}\right)\right)^{2}+k^{2}_{\check{y}_{i}},
\label{53}
\end{eqnarray}
and
\begin{eqnarray}
&\displaystyle 
\tan \delta_{i}\left(t\right)=k_{\check{y}_{i}}\left(k_{\check{x}_{i}}e^{\bar{U}'_{ix}t}
-k_{\check{y}_{i}}\left(V'_{0}t+\frac{\bar{U}_{iy}'}{\bar{U}_{ix}'}\right)\right)^{-1}.
\label{54}
\end{eqnarray}
The solution (\ref{50}) with potential (\ref{52}) is
\begin{eqnarray}
&\displaystyle 
\bar{f}_{i}\left(\tilde{v}_{i\perp}, \phi_{1}, v_{iz}, \check{\xi}_{i}, \check{\eta}_{i}, 
z_{i1}, t\right)=i\frac{e_{i}}{m_{i}}\int\limits^{t}_{0}dt_{1}
\nonumber 
\\ &\displaystyle
\times \int dk_{\check{x}_{i}}dk_{\check{y}_{i}}dk_{z}\Phi_{i}\left(k_{\check{x}_{i}}, 
k_{\check{y}_{i}}, k_{z}, t_{1}\right)
\nonumber 
\\ &\displaystyle
\times
e^{i\left(k_{\check{x}_{i}}\check{\xi}_{i}+k_{\check{y}_{i}}\check{\eta}_{i}+k_{z}z_{i}\right)}
\nonumber 
\\ &\displaystyle 
\times \sum\limits_{n=-\infty}^{\infty}J_{n}\left(\frac{k_{i\bot}\left(t_{1}\right)
\tilde{v}_{i\perp}}{\omega_{ci}}\right)e^{-in\left(\phi_{1}-
\omega_{ci}t_{1}-\delta\left(t_{1}\right)\right)}
\nonumber 
\\ &\displaystyle
\times
\left(\frac{k_{\check{y}_{i}}}{\omega_{ci}}e^{\bar{U}'_{ix}t_{1}}
\frac{\partial \bar{F}_{i0}}{\partial \check{\xi}_{i}}+\frac{n\omega_{ci}}{\tilde{v}_{i\perp}}
\frac{\partial \bar{F}_{i0}}{\partial 
\tilde{v}_{i\perp}}+k_{z}\frac{\partial \bar{F}_{i0}}{\partial v_{iz}}\right).
\label{55}
\end{eqnarray}
In what follows, we consider the stability of the microscale perturbations of the convective 
flows with wavelength much less than the plasma inhomogeneity scale length $L_{\check{X}_{i}}$, 
for which $|k_{i\bot}L_{\check{x}_{i}}|\gg 1$ and the Fourier transform of $\bar{f}_{i}$ over $
\check{x}_{i}$ can be performed in the local approximation. The Fourier transformed microscale
perturbation of the ion density, determined within this approximation, is 
\begin{widetext}
\begin{eqnarray}
&\displaystyle 
n_{i}\left(k_{\check{x}_{i}}, k_{\check{y}_{i}}, k_{z}, t\right)
=i\frac{2\pi e_{i}}{m_{i}}\sum\limits_{n=-\infty}^{\infty}\int\limits^{t}_{0}dt_{1}
\Phi_{i}\left(k_{\check{x}_{i}}, k_{\check{y}_{i}}, k_{z}, t_{1}\right)
\int\limits^{\infty}_{-\infty}dv_{iz}\int\limits_{0}^{\infty}d\tilde{v}_{i\perp}\tilde{v}_{i\perp}
\nonumber 
\\ &\displaystyle
\times\sum\limits_{n=-\infty}^{\infty}J_{n}\left(\frac{k_{i\bot}\left(t\right)
\tilde{v}_{i\perp}}{\omega_{ci}}\right)J_{n}\left(\frac{k_{i\bot}\left(t_{1}\right)
\tilde{v}_{i\perp}}{\omega_{ci}}\right)e^{-ik_{z}v_{iz}\left(t-t_{1}\right)
-in\left(\omega_{ci}\left(t-t_{1}\right)-\delta\left(t\right)
+\delta\left(t_{1}\right)\right)}
\nonumber 
\\ &\displaystyle
\times
\left(\frac{k_{\check{y}_{i}}}{\omega_{ci}}e^{\bar{U}'_{ix}t}
\frac{\partial \bar{F}_{i0}}{\partial \check{\xi}_{i}}+\frac{n\omega_{ci}}{\tilde{v}_{i\perp}}
\frac{\partial \bar{F}_{i0}}{\partial 
\tilde{v}_{i\perp}}+k_{z}\frac{\partial \bar{F}_{i0}}{\partial v_{iz}}\right).
\label{56}
\end{eqnarray}
\end{widetext}
For the Maxwellian distribution $\bar{F}_{i0}\left(\tilde{\mathbf{v}}_{i}, 
\check{\xi}_{i}\right)$ of ions with initial value (\ref{46}) in the case of the uniform ion 
temperature Eq. (\ref{56}) gives
\begin{eqnarray}
&\displaystyle 
n_{i}\left(k_{\check{x}_{i}}, k_{\check{y}_{i}}, k_{z}, t\right)
=i\frac{n_{0i}\left(\check{\xi}\right)e_{i}}{T_{i}}
\nonumber 
\\ &\displaystyle
\times \sum\limits_{n=-\infty}^{\infty}
\int\limits^{t}_{0}dt_{1}\Phi_{i}\left(k_{\check{x}_{i}}, k_{\check{y}_{i}}, k_{z}, t_{1}\right)
\nonumber 
\\ &\displaystyle
\times I_{n}\left(k_{i\bot}\left(t\right)k_{i\bot}\left(t_{1}\right)\rho^{2}_{i}
\right)e^{-\frac{1}{2}\rho^{2}_{i}\left(k_{i\bot}^{2}\left(t\right)+k_{i\bot}^{2}
\left(t_{1}\right)\right)}
\nonumber 
\\ &\displaystyle
\times e^{-\frac{1}{2}k^{2}_{z}v_{z}\left(t-t_{1}\right)^{2}
-in\left(\omega_{ci}\left(t-t_{1}\right)-\delta\left(t\right)
+\delta\left(t_{1}\right)\right)}
\nonumber 
\\ &\displaystyle
\times
\left(k_{\check{y}_{i}}v_{di}e^{\bar{U}'_{ix}t}
-n\omega_{ci}+ik^{2}_{z}v^{2}_{Ti}\left(t-t_{1}\right)\right).
\label{57}
\end{eqnarray}
where $v_{d\alpha}=(cT_{\alpha}/e_{\alpha}B_{0})d\ln n_{0}(\check{x}_{i})/d\check{x}_{i}$ 
is the ion $\left(\alpha=i\right)$ and electron $\left(\alpha=e\right)$ diamagnetic velocity,
$\rho_{i}$ is the thermal ion Larmor radius, and $I_{n}$ is the modified Bessel function 
of the first kind and order $n$.
The Fourier transform $n_{e}\left(k_{\check{x}_{e}}, k_{\check{y}_{e}}, k_{z}, t\right)$ 
of the microscale  perturbation of the electron density, performed in the electron frame with 
coordinates $\check{x}_{e}, \check{y}_{e}, z_{e1}, t$, is determined in the same way as it is
given by Eqs. (\ref{34})-(\ref{58}) for $n_{i}\left(k_{\check{x}_{i}}, k_{\check{y}_{i}}, 
k_{z}, t\right)$ with changed ion on electron species subscripts. The derived ion 
and electron density perturbations are employed in the Poisson equation, Fourier transformed 
over the variables $\check{x}_{i}, \check{y}_{i}$. Therefore $n_{e}\left(k_{\check{x}_{e}}, 
k_{\check{y}_{e}}, k_{z}, t\right)$ for the Poisson equation should be recalculated in the 
variables $\check{x}_{i}, \check{y}_{i}$ of the ion frame. For this goal, we derive 
the relations between variables $\check{x}_{i}, \check{y}_{i}$ and $\check{x}_{e}, 
\check{y}_{e}$. Because the difference between $\tilde{x}_{i}$ and $\tilde{x}_{e}$,  as well 
as between $\tilde{y}_{i}$ and $\tilde{y}_{e}$, are on the order of the microscale displacements 
of the ions relative to electrons in FW field, which are on the order of or less than the wavelength 
of the microscale perturbations, we can use relations $\tilde{x}_{i}=\tilde{x}_{e}$,  
and $\tilde{y}_{i}=\tilde{y}_{e}$ with Eqs. (\ref{40}), (\ref{41}) and obtain on this way the
relations
\begin{widetext}
\begin{eqnarray}
&\displaystyle \check{x}_{e}\left(\check{x}_{i}, t\right)=\frac{1}{\bar{U}'_{ex}}\left\lbrace 
e^{\bar{U}'_{ex}t}
\left[\bar{U}^{(0)}_{ex}+\frac{\bar{U}'_{ex}}{{\bar{U}'_{ix}}}\left[\left(\bar{U}^{(0)}_{ix}+
\bar{U}'_{ix}\check{x}_{i}\right)e^{-\bar{U}'_{ix}t}- \bar{U}^{(0)}_{ix} \right] \right]
-\bar{U}^{(0)}_{ex}\right\rbrace,
\label{58}
\end{eqnarray}
\begin{eqnarray}
&\displaystyle \check{y}_{e}\left(\check{y}_{i}, \check{x}_{i}, t\right)=\check{y}_{i}-\left[ 
\left(\bar{U}^{(0)}_{iy}
-\bar{U}^{(0)}_{ix}\frac{\bar{U}'_{iy}}{{\bar{U}'_{ix}}}\right)
- \left(\bar{U}^{(0)}_{ey}-\bar{U}^{(0)}_{ex}\frac{\bar{U}'_{ey}}{{\bar{U}'_{ex}}}\right)\right]t
\nonumber 
\\ &\displaystyle
+\frac{\bar{U}^{(0)}_{ix}}{\bar{U}'_{ix}}\left(\frac{\bar{U}'_{iy}}{\bar{U}'_{ix}}-
\frac{\bar{U}'_{ey}}{\bar{U}'_{ex}}\right)e^{-\bar{U}'_{ix}t}
+\frac{\bar{U}'_{ey}}{\bar{U}'_{ex}}\left(\frac{\bar{U}^{(0)}_{ix}}{\bar{U}'_{ix}}-
\frac{\bar{U}^{(0)}_{ex}}{\bar{U}'_{ex}}
\right)
+\left(\frac{\bar{U}'_{iy}}{\bar{U}'_{ix}}-\frac{\bar{U}'_{ey}}{\bar{U}'_{ex}}\right)
\check{x}_{i}e^{-\bar{U}'_{ix}t}.
\label{59}
\end{eqnarray}
\end{widetext}
Note, the relations for $\check{x}_{i}\left(\check{x}_{e}, t\right)$ and for $\check{y}_{i}
\left(\check{y}_{e}, \check{x}_{e}, t\right)$ are derived by 
changing species subscripts $i\leftrightarrows e$ in Eqs. (\ref{58}), (\ref{59}). 

The equation for $n_{e}\left(k_{\check{x}_{e}}, k_{\check{y}_{e}}, k_{z}, t\right)$, 
similar to (\ref{56}) for $n_{i}$, contains the Fourier transform $\Phi_{e}\left(k_{\check{x}_{e}}, 
k_{\check{y}_{e}}, k_{z}, t_{1}\right)$ of the potential $\Phi_{e}\left(\check{x}_{e}, 
\check{y}_{e}, z, t_{1}\right)$. The connection relation of $\Phi_{e}\left(k_{\check{x}_{e}}, 
k_{\check{y}_{e}}, k_{z}, t_{1}\right)$ with $\Phi_{i}\left(k_{\check{x}_{i}}, 
k_{\check{y}_{i}}, k_{z}, t_{1}\right)$ follows from the relation
\begin{widetext}
\begin{eqnarray}
&\displaystyle 
\Phi_{e}\left(k_{\check{x}_{e}}, k_{\check{y}_{e}}, k_{z}, t_{1}\right)=
\int d\check{x}_{e}\int d\check{y}_{e}\Phi\left(\check{x}_{e}, \check{y}_{e}, k_{z}, 
t_{1}\right)e^{-i\left(k_{\check{x}_{e}}\check{x}_{e}+k_{\check{y}_{e}}\check{y}_{e}\right)} 
\nonumber 
\\ &\displaystyle
=\frac{1}{\left(2\pi\right)^{2}}\int dk_{\check{x}_{i}}\int dk_{\check{y}_{i}}\Phi_{i}\left(k_{\check{x}_{i}}, k_{\check{Y}_{i}}, k_{z}, t_{1}\right)
\int d\check{x}_{i}\int d\check{y}_{i}
\frac{\partial\left(\check{x}_{e}, \check{y}_{e} \right)}
{\partial\left(\check{x}_{i}, \check{y}_{i}\right)}
\nonumber 
\\ &\displaystyle
\times e^{i\left(k_{\check{x}_{i}} - k_{\check{x}_{e}}\right)\check{x}_{i}+ 
i\left(k_{\check{y}_{i}}-k_{\check{y}_{e}}\right)\check{y}_{i}
-ik_{\check{x}_{e}}\left(\check{x}_{e}-\check{x}_{i}\right)
-ik_{\check{y}_{e}}\left(\check{y}_{e}-\check{y}_{i}\right)}
\nonumber 
\\ &\displaystyle
=\Phi_{i}\left(k_{\check{x}_{e}}
+k_{\check{x}_{e}}b_{1x}\left(t_{1}\right)+k_{\check{y}_{e}}b_{1y}\left(t_{1}\right), 
k_{\check{y}_{e}}, k_{z},  t_{1}\right)
e^{\left(\bar{U}'_{ex}-\bar{U}'_{ix}\right)t_{1}}
e^{-ik_{\check{x}_{e}}b_{0x}
\left(t_{1}\right)-ik_{\check{y}_{e}}b_{0y}\left(t_{1}\right)}.
\label{60}
\end{eqnarray}
\end{widetext}
In Eq. (\ref{60}), 
\begin{eqnarray}
&\displaystyle 
\frac{\partial\left(\check{x}_{e}, \check{y}_{e}\right)}{\partial\left(\check{x}_{i}, 
\check{y}_{i}\right)}=e^{\left(\bar{U}'_{ex}-\bar{U}'_{ix}\right)t_{1}}
\label{61}
\end{eqnarray}
is the Jacobian of the transformation $\check{x}_{e}, \check{y}_{e}$ to $\check{x}_{i}, 
\check{y}_{i}$, and the relations 
\begin{eqnarray}
&\displaystyle 
\check{x}_{e}\left(\check{x}_{i}, t_{1}\right)-\check{x}_{i}=b_{0x}\left(t_{1}\right)
+b_{1x}\left(t_{1}\right)\check{x}_{i}
\label{62}
\end{eqnarray}
and 
\begin{eqnarray}
&\displaystyle 
\check{y}_{e}\left(\check{y}_{i}, \check{x}_{i}, t_{1}\right)-\check{y}_{i}=b_{0y}\left(t_{1}\right)
+b_{1y}\left(t_{1}\right)\check{x}_{i},
\label{63}
\end{eqnarray}
where
\begin{eqnarray}
&\displaystyle 
b_{0x}\left(t_{1}\right)=\frac{1}{\bar{U}'_{ex}}
e^{\bar{U}'_{ex}t_{1}}
\left[\bar{U}^{(0)}_{ex}+\bar{U}^{(0)}_{ix}\frac{\bar{U}'_{ex}}{{\bar{U}'_{ix}}}\left(e^{-
\bar{U}'_{ix}t_{1}}-1\right)\right]
\nonumber 
\\ &\displaystyle
-\frac{\bar{U}^{(0)}_{ix}}{{\bar{U}'_{ix}}},
\label{64}
\end{eqnarray}
\begin{eqnarray}
&\displaystyle 
b_{1x}\left(t_{1}\right)=\left(e^{\left(\bar{U}'_{ex}-\bar{U}'_{ix}\right)t_{1}}-1\right),
\label{65}
\end{eqnarray}
\begin{eqnarray}
&\displaystyle 
b_{0y}\left(t_{1}\right)=\left[ 
\left(\bar{U}^{(0)}_{ey}
-\bar{U}^{(0)}_{ex}\frac{\bar{U}'_{ey}}{{\bar{U}'_{ex}}}\right)
- \left(\bar{U}^{(0)}_{iy}-\bar{U}^{(0)}_{ix}\frac{\bar{U}'_{iy}}{{\bar{U}'_{ix}}}\right)
\right]t_{1}
\nonumber 
\\ &\displaystyle
+\frac{\bar{U}^{(0)}_{ix}}{\bar{U}'_{ix}}\left(\frac{\bar{U}'_{iy}}{\bar{U}'_{ix}}-
\frac{\bar{U}'_{ey}}{\bar{U}'_{ex}}\right)e^{-\bar{U}'_{ix}t_{1}}
\nonumber 
\\ &\displaystyle
+\frac{\bar{U}'_{ey}}{\bar{U}'_{ex}}\left(\frac{\bar{U}^{(0)}_{ix}}{\bar{U}'_{ix}}-
\frac{\bar{U}^{(0)}_{ex}}{\bar{U}'_{ex}}
\right),
\label{66}
\end{eqnarray}
\begin{eqnarray}
&\displaystyle 
b_{1y}\left(t_{1}\right)=\left(\frac{\bar{U}'_{iy}}{\bar{U}'_{ix}}-
\frac{\bar{U}'_{ey}}{\bar{U}'_{ex}}\right)e^{-\bar{U}'_{ix}t_{1}},
\label{67}
\end{eqnarray}
were used.

Now we determine the relation between the Fourier transform $n_{e}\left(k_{\check{x}_{e}}, 
k_{\check{y}_{e}}, k_{z}, t\right)$ of the electron density perturbation 
$n_{e}\left(\check{x}_{e}, \check{y}_{e}, z_{e}, t\right)$, performed in the electron 
frame with 
variables $\check{x}_{e}, \check{y}_{e}$, with the Fourier transform $n^{(i)}_{e}
\left(k_{\check{x}
_{i}}, k_{\check{y}_{i}}, k_{z}, t\right)$ of $n_{e}\left(\check{x}_{e}, \check{y}_{e}, Z_{e}, t
\right)$, performed in the ion frame with variables $\check{x}_{i}, \check{y}_{i}$. 
\begin{widetext}
\begin{eqnarray}
&\displaystyle 
n^{(i)}_{e}\left(k_{\check{x}_{i}}, k_{\check{y}_{i}}, k_{z}, t\right)
=\int d\check{x}_{i} \int d\check{y}_{i}
e^{-i\left(k_{\check{x}_{i}}\check{x}_{i}
+k_{\check{y}_{i}}\check{y}_{i}\right)}n_{e}\left(\check{x}_{e}, 
\check{y}_{e}, k_{z}, t\right)
\nonumber 
\\ &\displaystyle
=\int d\check{x}_{e} \int d\check{y}_{e}n_{e}\left(\check{x}_{e}, \check{y}_{e}, k_{z}, 
t\right)\frac{\partial\left(\check{x}_{i}, \check{y}_{i}\right)}{\partial\left(\check{x}_{e}, 
\check{y}_{e}\right)}e^{-ik_{\check{x}_{i}}\check{x}_{e}
-ik_{\check{y}_{i}}\check{y}_{e}-ik_{\check{x}_{i}}\left(\check{x}_{i}-\check{x}_{e}\right)
-ik_{\check{y}_{i}}\left(\check{y}_{i}-\check{y}_{e}\right)}
\nonumber 
\\ &\displaystyle
=\int d\check{x}_{e} \int d\check{y}_{e}n_{e}\left(\check{x}_{e}, \check{y}_{e}, 
k_{z}, t\right)e^{\left(\bar{U}'_{ix}-\bar{U}'_{ex}\right)t}
\nonumber 
\\ &\displaystyle
\times e^{-ik_{\check{x}_{i}}\check{x}_{e}
-ik_{\check{y}_{i}}\check{y}_{e}-ik_{\check{x}_{i}}\left(a_{0x}\left(t\right)
+a_{1x}\left(t\right)\check{x}_{e}\right)-ik_{\check{y}_{i}}
\left(a_{0y}\left(t\right)+a_{1y}\left(t\right)\check{x}_{e}\right)}
\nonumber 
\\ &\displaystyle
=e^{\left(\bar{U}'_{ix}-\bar{U}'_{ex}\right)t}e^{-ik_{\check{x}_{i}}
a_{0x}\left(t\right)-ik_{\check{y}_{i}}a_{0y}\left(t\right)}n_{e}
\left(k_{\check{x}_{i}}\left(1+a_{1x}\left(t\right)\right)
+k_{\check{y}_{i}}a_{1y}\left(t\right), k_{\check{y}_{i}}, k_{z}, t\right),
\label{68}
\end{eqnarray}
\end{widetext}
where the relations
\begin{eqnarray}
&\displaystyle 
\check{x}_{i}\left(\check{x}_{e}, t\right)-\check{x}_{e}=a_{0x}\left(t\right)
+a_{1x}\left(t\right)\check{x}_{e}
\label{69}
\end{eqnarray}
and 
\begin{eqnarray}
&\displaystyle 
\check{y}_{i}\left(\check{y}_{e}, \check{x}_{e}, t\right)-\check{y}_{e}=a_{0y}\left(t\right)
+a_{1y}\left(t\right)\check{x}_{e},
\label{70}
\end{eqnarray}
where used. The functions $a_{0x}\left(t\right)$, $a_{1x}\left(t\right)$, $a_{0y}\left(t\right)$, 
and $a_{1y}\left(t\right)$ are determined by the functions $b_{0x}\left(t\right)$, 
$b_{1x}\left(t\right)$, $b_{0y}\left(t\right)$, and $b_{1y}\left(t\right)$, respectively, 
by changing species subscripts $i\leftrightarrows e$ in Eqs. (\ref{59}) -(\ref{64}).

By replacing $k_{\check{x}_{e}}$ and $k_{\check{y}_{e}}$ in Eq. (\ref{64}) on $k_{\check{x}_{i}}
\left(1+a_{1x}\left(t\right)\right) +k_{\check{y}_{i}}a_{1y}\left(t\right)$ 
and $k_{\check{y}_{i}}$, which, as it follows from Eq. (\ref{64}), are 
the new wave numbers conjugate with coordinates $\check{x}_{e}$ and 
$\check{y}_{e}$ in $n^{(i)}_{e}\left(k_{\check{x}_{i}}, k_{\check{y}_{i}}, k_{z}, t\right)$, 
we derive the following relation for $n^{(i)}_{e}$:
\begin{widetext}
\begin{eqnarray}
&\displaystyle 
n^{(i)}_{e}\left(k_{\check{x}_{i}}, k_{\check{y}_{i}}, k_{z}, t\right)=\frac{2i\pi e}{m_{e}}
e^{-ik_{\check{x}_{i}}a_{0x}\left(t\right)-ik_{\check{y}_{i}}a_{0y}\left(t\right)}
\int\limits^{t}_{0} dt_{1}e^{\left(\bar{U}'_{ix}-\bar{U}'_{ex}\right)\left(t-t_{1}\right)}
\nonumber 
\\ &\displaystyle
\times\Phi_{i}\left(\left(k_{\check{x}_{i}}
\left(1+a_{1x}\left(t\right)\right) +k_{\check{y}_{i}}
a_{1y}\left(t_{1}\right)\right)\left(1+b_{1x}\left(t_{1}\right)\right)+k_{\check{y}_{i}}b_{1y}
\left(t_{1}\right), k_{\check{y}_{i}}, k_{z},  t_{1}\right)
\nonumber 
\\ &\displaystyle
\times
e^{-i\left(k_{\check{x}_{i}}\left(1+a_{1x}\left(t\right)\right)+k_{\check{y}_{i}}
a_{1y}\left(t_{1}\right)\right)b_{0x}\left(t_{1}\right)-ik_{\check{y}_{i}}b_{0y}
\left(t_{1}\right)}
\nonumber 
\\ &\displaystyle
\times\int\limits^{\infty}_{0}dv_{e\bot}v_{e\bot}\int\limits_{-\infty}^{\infty}dv_{ez}
e^{-ik_{z}v_{ez}\left(t-t_{1}\right)}\left(\frac{k_{\check{y}_{i}}}{\omega_{ce}}
e^{\bar{U}'_{ex}t_{1}}\frac{\partial \bar{F}_{e}}{\partial \check{\xi}_{e}}
+k_{z}\frac{\partial \bar{F}_{e}}{\partial v_{ez}} \right).
\label{71}
\end{eqnarray}
\end{widetext}
Equation (\ref{71}) is valid for the perturbations with frequency much less than the electron 
cyclotron frequency and with wavelength across the magnetic field much larger than the thermal 
electron Larmor radius. 

The Poisson equation (\ref{49}) for $\Phi_{i}$, Fourier transformed over $\check{x}_{i}$ 
and $\check{y}_{i}$,
\begin{widetext}
\begin{eqnarray}
&\displaystyle 
\left(k^{2}_{\check{x}_{i}} e^{2\bar{U}_{ix}'t}-2e^{\bar{U}_{ix}'t}\left(V'_{0}t+\frac{\bar{U}_{iy}'}{\bar{U}_{ix}'}\right)k_{\check{x}_{i}}k_{\check{y}_{i}}
+\left(1+\left(V'_{0}t+\frac{\bar{U}_{iy}'}{\bar{U}_{ix}'}\right)^{2}\right)k^{2}_{\check{y}_{i}}
+k^{2}_{z}\right)\Phi_{i}\left(k_{\check{x}_{i}}, 
k_{\check{y}_{i}}, k_{z}, t\right)
\nonumber 
\\ &\displaystyle
=4\pi\left(e_{i}n_{i}\left(k_{\check{x}_{i}}, k_{\check{y}_{i}}, k_{z}, t
\right)-|e|n^{(i)}_{e}\left(k_{\check{x}_{e}}\left(k_{\check{x}_{i}}, k_{\check{y}_{i}}, t\right), 
k_{\check{y}_{i}}, k_{z}, t\right)\right),
\label{72}
\end{eqnarray}  
\end{widetext}
where $n_{i}$ and $n^{(i)}_{e}$ are determined by Eqs. (\ref{54}) and (\ref{71}) respectively,
is the equation which determines the temporal evolution of the single spatial Fourier mode 
$\Phi_{i}\left(k_{\check{x}_{i}}, k_{\check{y}_{i}}, k_{z}, t\right)$
in the compressed-sheared flow. 

Now we consider the particular cases for Eq.(\ref{72}), in which $n_{i}$ and $n^{(i)}_{e}$ are 
determined by Eqs. (\ref{57}) and (\ref{71}). 

1. In the case of the currentless compressed-sheared flow $\bar{U}^{(0)}_{ix}= 
\bar{U}^{(0)}_{ex}$, $\bar{U}_{ix}'= \bar{U}_{ex}'$, and $\bar{U}^{(0)}_{iy}= 
\bar{U}^{(0)}_{ey}$, $\bar{U}_{iy}'= \bar{U}_{ey}'$. It follows from Eqs. (\ref{62})-(\ref{67})
that in this case $b_{0x}\left(t_{1}\right)=a_{0x}\left(t\right)=0$, $b_{1x}\left(t_{1}\right)
=a_{1x}\left(t\right)=0$, and $b_{0y}\left(t_{1}\right)=a_{0y}\left(t\right)=0$, 
$b_{1y}\left(t_{1}\right)=a_{1y}\left(t\right)=0$. Therefore in this case 
$\check{x}_{i}=\check{x}_{e}$,
$\check{y}_{i}=\check{y}_{e}$, and $\Phi_{e}\left(k_{\check{x}_{e}}, k_{\check{y}
_{e}}, k_{z}, t\right)=\Phi_{i}\left(k_{\check{x}_{i}}, k_{\check{y}_{i}}, k_{z}, t\right)$ and 
$n^{(i)}_{e}\left(k_{\check{x}_{i}}, k_{\check{y}_{i}}, 
k_{z}, t\right)=n_{e}\left(k_{\check{x}_{e}}, k_{\check{y}_{e}}, k_{z}, t\right)$. Equation 
(\ref{72}) in this case has a form

\begin{widetext}
\begin{eqnarray}
&\displaystyle 
\lambda^{2}_{Di}\left(k^{2}_{i\bot}\left(t\right)+k^{2}_{z}\right)\Phi_{i}\left(k_{\check{x}_{i}}, 
k_{\check{y}_{i}}, k_{z}, t
\right)=\sum\limits_{n=-\infty}^{\infty}
\int\limits^{t}_{t_{0}}dt_{1}\Phi_{i}\left(k_{\check{x}_{i}}, k_{\check{y}_{i}}, k_{z}, t_{1}\right)
I_{n}\left(k_{i\bot}\left(t\right)k_{i\bot}
\left(t_{1}\right)\rho^{2}_{i}\right)
\nonumber 
\\ &\displaystyle
\times
e^{-\frac{1}{2}\rho^{2}_{i}\left(k_{i\bot}^{2}\left(t\right)+k_{i\bot}^{2}
\left(t_{1}\right)\right)}\left(ik_{\check{y}_{i}}v_{di}e^{\bar{U}'_{ix}t_{1}}
-in\omega_{ci}-k^{2}_{z}v^{2}_{Ti}\left(t-t_{1}\right)\right)e^{-\frac{1}{2}k^{2}_{z}v^{2}_{Ti}
\left(t-t_{1}\right)^{2}
-in\left(\omega_{ci}\left(t-t_{1}\right)-\delta\left(t\right)
+\delta\left(t_{1}\right)\right)}
\nonumber 
\\ &\displaystyle
=\frac{T_{i}}{T_{e}}\int\limits^{t}_{t_{0}} dt_{1}\Phi_{i}\left(k_{\check{x}
_{i}}, k_{\check{y}_{i}}, k_{z}, t_{1}\right)e^{-\frac{1}{2}k^{2}_{z}v^{2}_{Te}
\left(t-t_{1}\right)^{2}} 
\left[ik_{\check{y}_{i}}v_{de}e^{\bar{U}'_{ex}t_{1}}
-k^{2}_{z}v^{2}_{Te}\left(t-t_{1}\right)\right].
\label{73}
\end{eqnarray}
\end{widetext}
where $t_{0}\geq 0$, $\lambda_{Di(e)}$ is the ion (electron) Debye length, and 
\begin{eqnarray}
&\displaystyle 
A_{in}\left(t, t_{1}\right)=I_{n}\left(k_{i\bot}\left(t\right)k_{i\bot}
\left(t_{1}\right)\rho^{2}_{i}\right)
\nonumber 
\\ &\displaystyle
\times
e^{-\frac{1}{2}\rho^{2}_{i}\left(k_{i\bot}^{2}\left(t\right)+k_{i\bot}^{2}
\left(t_{1}\right)\right)}.
\label{74}
\end{eqnarray}
Equation (\ref{73}) was derived for the first time in Ref.\cite{Mikhailenko4} 
for the poloidal sheared plasma flow
without the convective flows (i. e. for the case $\bar{U}'_{ix}=\bar{U}'_{ex}=0$ and 
$\bar{U}'_{iy}=\bar{U}'_{ey}=0$). In that case, the poloidal velocity shear manifests as a time-
dependence of $A_{in}\left(t, t_{1}\right)$ function which determines effect of the finite ion Larmor 
radius. The solution of Eq. (\ref{73}), derived for the kinetic drift instability in the poloidal 
sheared flow, displays\cite{Mikhailenko4} the nonmodal effect of the reduction with time of the 
frequency and of the growth rate of this instability caused by the flow velocity shear. By the integration by parts of the first term on the right part 
of Eq. (\ref{73}), this equation for the low frequency perturbations, for which $d\Phi/dt\ll \omega_{ci}\Phi$,   may be presented in the form  similar to Eq. (25) of Ref. \cite{Mikhailenko4}, 
\begin{widetext}
\begin{eqnarray}
&\displaystyle 
\int\limits_{t{0}}^{t} dt_{1}\frac{d}{dt_{1}}\left[\Phi_{i}\left(k_{\check{x}_{i}}, 
k_{\check{y}_{i}}, k_{z}, t_{1}\right)\left(1+\frac{T_{i}}{T_{e}}-A_{in}\left(t, 
t_{1}\right)\right)\right] 
\nonumber 
\\ &\displaystyle
-i\int\limits^{t}_{t_{0}}dt_{1}\Phi_{i}k_{\check{y}_{i}}v_{di}e^{\bar{U}'_{ex}t_{1}}
A_{in}\left(t, t_{1}\right)
\nonumber 
\\ &\displaystyle
=\frac{T_{i}}{T_{e}}\int\limits^{t}_{t_{0}}\left(\frac{d\Phi_{i}}{dt_{1}}+ik_{\check{y}_{i}}v_{de}
e^{\bar{U}'_{ex}t_{1}}\Phi_{i}\right)e^{-\frac{1}{2}k^{2}_{z}v^{2}_{Te}
\left(t-t_{1}\right)^{2}}
\label{75}
\end{eqnarray}
\end{widetext}
We found that in the time domain $\left(t, t_{0}\right)$, in which $k_{i\bot}\left(t_{1}\right)
\rho_{i}\gg 1$, the temporal evolution of the potential $\Phi_{i}$ in the compressed flow, 
predicted by the solution to Eq. (\ref{75}), resembles the temporal evolution of the potential 
$\Phi_{i}$ in the poloidal sheared flow: the potential $\Phi_{i}$ gradually becomes a zero-
frequency cell-like perturbation when time elapsed. 

 \section{Conclusions}\label{sec5} 
 A nonmodal kinetic theory of the stability of the two-dimensional 
compressed-sheared mesoscale plasma flows, generated by the radially inhomogeneous electrostatic 
ion cyclotron parametric microturbulence in the pedestal plasma with a sheared poloidal flow, is 
developed. This theory reveals that the separate spatially uniform Fourier modes of the 
electrostatic responses of the ions and of the electrons on the mesoscale convective flows 
are determined only in the 
frames of references moved with velocities of the ion and electron convective flows. 
In the laboratory frame, these modes are observed as the compressed-sheared modes with time 
dependent wave numbers. The integral equation, which governs the separate Fourier mode of the 
electrostatic potential of the plasma species responses on the mesoscale convective flows, 
is derived. In this equation, the effects of the compressing and shearing of the convective flows 
are revealed as the time dependence of the finite ion Larmor radius effect. The solution of this 
equation for the kinetic drift instability displays the nonmodal transformation of the 
potential to the zero frequency cell-like perturbation when time elapsed. 
 
 \begin{acknowledgments}
This work was supported by National R\&D Program through the National Research Foundation of 
Korea (NRF) funded by the Ministry of Education, Science and Technology (Grant No. 
NRF-2018R1D1A3B07051247) and BK21 FOUR, 
the Creative Human Resource Education and Research Programs for ICT Convergence in the 4th 
Industrial Revolution.
\end{acknowledgments}

\bigskip
{\bf DATA AVAILABILITY}

\bigskip
The data that support the findings of this study are available from the corresponding author upon 
reasonable request.

\appendix 
\section{Velocities $\tilde{U}_{ix}\left(\tilde{X}_{i}\right)$ and 
$\tilde{U}_{iy}\left(\tilde{X}_{i}\right)$ of the convective flows}
\label{sec6} 

For the electric field $\tilde{\mathbf{E}}_{i}$, given by Eq. (\ref{8}), velocities 
$\bar{U}_{ix}\left(\tilde{X}_{i}\right)$ and $\bar{U}_{iy}
\left(\tilde{X}_{i}\right)$, after long calculation similar to performed in 
Ref.\cite{Mikhailenko1}, are determined by the following relations: 
\begin{widetext}
\begin{eqnarray}
&\displaystyle 
\bar{U}_{ix}\left(\tilde{X}_{i}\right)= \frac{1}{2\omega_{ci}}\frac{e_{i}^{2}}
{m^{2}_{i}}\frac{1}{4\pi^{2}}\sum_{n}
\int d\mathbf{k}\left[a_{i1}\left(\mathbf{k}, n\right)\tilde{E}_{ix}
\left(\mathbf{k}, \tilde{X}_{i}, n\right)
\frac{\partial}{\partial \tilde{X}_{i}}\left(\tilde{E}^{\ast}_{iy}
\left(\mathbf{k}, \tilde{X}_{i}, n\right)\right)\right.
\nonumber\\ 
& \displaystyle
\left.+a_{i2}\left(\mathbf{k}, n\right)\tilde{E}_{iy}\left(\mathbf{k}, \tilde{X}_{i}, n\right)
\frac{\partial}{\partial \tilde{X}_{i}}
\left(\tilde{E}^{\ast}_{ix}\left(\mathbf{k},\tilde{X}_{i}, n\right)\right)\right]
\label{81}
\end{eqnarray}
and
\begin{eqnarray}
&\displaystyle 
\bar{U}^{(0)}_{iy}\left(\tilde{X}_{i}\right)= -\frac{1}{4\omega_{ci}}\frac{e_{i}^{2}}
{m^{2}_{i}}\frac{1}{4\pi^{2}}\sum_{n}\int d\mathbf{k}\left[ a_{i1}\left(\mathbf{k}, n\right)\frac{\partial}
{\partial \tilde{X}_{i}}\left|\tilde{E}_{ix}\left(\mathbf{k}, \tilde{X}_{i}, n\right)\right|^{2}
-a_{i2}\left(\mathbf{k}, n\right)\frac{\partial}{\partial X_{i}}
\left|\tilde{E}_{iy}\left(\mathbf{k},\tilde{X}_{i}, n\right)\right|^{2}\right]
\label{82}
\end{eqnarray}
where the asterisk in Eq. (\ref{81}) implies the operation of complex conjugate. The 
coefficients $a_{i1}\left(\mathbf{k}, n\right)$ 
and $a_{i2}\left(\mathbf{k}, n\right)$ are determined as
\begin{eqnarray}
&\displaystyle a_{i1}\left(\mathbf{k}, n\right)=\left[\frac{\omega_{ci}}
{\omega_{n}\left(\mathbf{k}\right)\left(\omega_{ci}+\omega_{n}\left(\mathbf{k}\right)
\right)^{2}} +\frac{\omega_{ci}}{\omega_{n}\left(\mathbf{k}
\right)\left(\omega_{ci}-\omega_{n}\left(\mathbf{k}\right)\right)^{2}}+
\frac{1}{\left(\omega^{2}_{ci}-\omega_{n}^{2}\left(\mathbf{k}\right)\right)}\right], 
\label{83}
\end{eqnarray}
and
\begin{eqnarray}
&\displaystyle a_{i2}\left(\mathbf{k}\right)=\left[\frac{1}{\left(\omega_{ci}
+\omega_{n}\left(\mathbf{k}
\right)\right)^{2}} +\frac{1}{\left(\omega_{ci}-\omega_{n}\left(\mathbf{k}\right)
\right)^{2}}+\frac{1}{\left(\omega^{2}_{ci}-\omega_{n}^{2}\left(\mathbf{k}\right)\right)}\right].
\label{84}
\end{eqnarray}
\end{widetext}
The velocities of electrons $\bar{U}_{ex}\left(\tilde{X}_{i}\right)$ 
and $\bar{U}_{ey}\left(\tilde{X}_{i}\right)$ in the ion frame, with  
electric field $\mathbf{E}_{e}\left(\mathbf{\hat{r}}_{i}, \tilde{X}_{i}, t\right)$ 
determined by Eq. (\ref{10}), are
\begin{eqnarray}
&\displaystyle \bar{U}_{ex}\left(\tilde{X}_{i}\right)\approx \frac{c^{2}}{B^{2}_{0}}
\frac{1}{4\pi^{2}}\sum_{n}\int d\mathbf{k}\tilde{E}_{ix}\left(\mathbf{k}, \tilde{X}_{i}, n\right)
\nonumber\\ 
& \displaystyle
\times\frac{\partial}{\partial \tilde{X}_{i}}\left(\tilde{E}^{\ast}_{iy}\left(\mathbf{k}, 
\tilde{X}_{i}, n\right)\right)\sum\limits_{p=-\infty}^{\infty}J^{2}_{p}\left(a_{ie}
\left(\mathbf{k}, \tilde{X}_{i}\right)\right)
\nonumber\\ 
& \displaystyle
\times
\frac{1}{\Omega_{p}\left(\mathbf{k}, \tilde{X}_{i}\right)},
\label{85}
\end{eqnarray}
and 
\begin{eqnarray}
&\displaystyle \bar{U}_{ey}\left(\tilde{X}_{i}\right)\approx 
-\frac{1}{2}\frac{c^{2}}{B^{2}_{0}}\frac{1}{4\pi^{2}}\sum_{n}
\int d\mathbf{k}\frac{\partial}{\partial \tilde{X}_{i}}\left|\tilde{E}_{ix}
\left(\mathbf{k}, \tilde{X}_{i}, n\right)\right|^{2}
\nonumber\\ 
& \displaystyle
\times\sum\limits_{p=-\infty}^{\infty}J^{2}_{p}\left(a_{ie}\left(\mathbf{k}, 
\tilde{X}_{i}\right)
\right)\frac{1}{\Omega_{p}\left(\mathbf{k}, \tilde{X}_{i}\right)},
\label{86}
\end{eqnarray}
where limit $|\omega_{ce}|\gg \Omega_{p}\sim \omega_{ci}$ was used.

\end{document}